\documentclass[useAMS,usenatbib]{mn2e}
\usepackage{graphicx}
\usepackage{amssymb}

\newcommand{\apj}{ApJ}

\newcommand{\aap}{A\&A}

\newcommand{\mnras}{MNRAS}

\newcommand{\MC}{\multicolumn}
\newcommand{\kms}{km\,s$^{-1}$}

\newcommand{\HI}{H{\sc i}}
\newcommand{\HII}{H{\sc ii}}
\newcommand{\HeI}{He{\sc i}}
\newcommand{\sunn}{$_{\odot}$}
\newcounter{qub}
\newcommand{\qq}{\addtocounter{qub}{1}\arabic{qub}}

\DeclareRobustCommand{\ion}[2]{%
\relax\ifmmode
\ifx\testbx\f
{\mathrm{#1\,\textsc{#2}}}\else
{\mathrm{#1\,\mathsc{#2}}}\fi
\else\textup{#1\,{\mdseries\textsc{#2}}}%
\fi}

\title[DDO~68: LBV, H$\alpha$ shells and the brightest stars]
{Extremely metal-poor galaxy DDO 68: the LBV, H$\alpha$ shells and
the most luminous stars}

\author[S.A. Pustilnik, L.N. Makarova,  Y.A. Perepelitsyna, A.V. Moiseev,
D.I. Makarov]{S.A. Pustilnik,$^1$\thanks{sap@sao.ru (SAP), lida@sao.ru (LNM),
moisav@sao.ru (AVM), dim@sao.ru (DIM)} L.N. Makarova,$^1$
Y.A. Perepelitsyna,$^1$ A.V. Moiseev,$^1$ D.I. Makarov$^1$  \\
$^1$ Special Astrophysical Observatory of RAS, Nizhnij Arkhyz,
Karachai-Circassia 369167, Russia}

\begin{document}

\label{firstpage}

\date{Accepted 2016 November 22. Received 2016 March 3}

\pagerange{\pageref{firstpage}--\pageref{lastpage}} \pubyear{2016}

\maketitle

\begin{abstract}

The paper presents  new results of the ongoing study of the unusual
Lynx-Cancer void galaxy DDO~68 with  record-low-metallicity regions
(12+$\log$(O/H)$\sim$7.14) of the current star formation (SF).
They include:
a) a new spectrum and photometry with the 6-m  SAO RAS telescope (BTA) for the
Luminous Blue Variable (LBV = DDO68-V1). Photometric data sets are
complemented with those based on the Sloan Digital Sky Survey (SDSS) and
the Hubble Space Telescope ({\it HST}) archive  images;
b) the analysis of the DDO~68 supergiant shell (SGS) and the prominent
smaller H$\alpha$ arcs/shells visible at the {\it HST} image coupled with
kinematics maps in H$\alpha$ obtained with the Fabry-Perot interferometer
(FPI) at the BTA;
c) the list of identified at the HST images of about 50  most luminous
stars ($ -9.1 < M_{\rm V} < -6.0$ mag) related to  star-forming
regions with the known extremely low O/H. This is intended to pave the path
for the actual science with the next generation of giant telescopes.
We confirm the earlier hints on  significant variations of the LBV
optical light deriving its amplitude of
$\Delta V \gtrsim$ 3.7~mag for the first time. New data suggest that in 2008--2010 the LBV
reached $M_{\rm V}$ = --10.5 and probably underwent a giant eruption.
We argue that the structure of star-forming complexes along the SGS
(`Northern Ring') perimeter provides evidence for the sequential induced SF episodes
caused by the shell gas instabilities and gravitational collapse.
The variability of some DDO~68 luminous extremely metal-poor stars
can be monitored with medium-size telescopes at sites with superb seeing.
\end{abstract}


\section[]{INTRODUCTION}
\label{sec:intro}

The dwarf irregular galaxy DDO~68 (UGC~5340, VV~542) resides in the nearby
Lynx-Cancer void \citep{PaperI}. It shows a peculiar, disturbed morphology
and hosts several prominent young star-forming (SF) regions with the record-low
metallicity (12+$\log$(O/H)=7.14).
Most of SF regions are
found at the periphery, mainly in the `Northern ring' and the `Southern tail'
(see \citet{DDO68}, hereafter PKP05, \citet{IT07}). In the repeated
observations of DDO~68, the unique Luminous Blue Variable star (LBV) was
discovered in one of the most metal-poor \HII\ regions in the local Universe
\citep{LBV,IT09}.

The GMRT (Giant Meterwave Radio Telescope) \HI-mapping study of DDO~68
revealed its complex structure and a velocity field consisting of two arms
winded asymmetrically around the main bright part of the galaxy's optical body.
Their properties do not contradict the assumption that the arms are of  tidal
origin formed as a result of the recent gas-rich merger \citep{Ekta08}.
Recent VLA (Very Large Array) and GBT (Green Bank Telescope) deep \HI\
maps revealed a faint companion (DDO~68C) at a projected distance of
$\sim$42~kpc from DDO~68. Its baryonic mass is $\sim$35 times smaller than
that of the main galaxy \citep{Cannon14}. There is a subtle \HI\ bridge
pulled-off of DDO~68C. However, due to a very large difference in mass and
a large distance, the reverse dynamical effect of DDO~68C to DDO~68
should be rather small \citep{Annibali16}.

In line with earlier results, the authors of the recent `qualitative'
analysis of DDO~68 based on the {\it HST} deep images in the broad filters F606W
and F814W \citep{Tikhonov14}  distinguish
two very different spatial components in the resolved stellar population with medium
($Z \sim Z$\sunn/5)
and very low ($Z \lesssim Z$\sunn/20) stellar metallicities. The first
relates to the `main, central' component, while the second one -- to the
`secondary, peripheric' component. They have also found that  the `secondary'
component (winded from the Northern periphery of the main body through the
Eastern edge to the Southern tail) has only a small fraction of old stars
which is not the case of the `main' component. In addition to this,
very deep imaging by \citet{Annibali16} reveals a much fainter low-surface
arc and a thin elongated fragment adjacent the main body, which the authors
assign to mini satellites being accreted to the `main' component.

The issue of nature of the secondary component remains unsettled.
In the recent paper by Makarov et al. (2016, MNRAS, in press), we show that
the small fraction of the secondary component's old population mentioned above
can be naturally explained for the `Northern ring' region
by the contribution of stars from the outer parts of the main,
central component. The latter fact implies that the small
fraction of old stars in this region does not belong to
the `secondary'
component of the assumed minor merger, with a measured {\it gas}
metallicity in its star-forming regions of $\sim Z$\sunn/35. Hence, it can be
really a very young object that formed its main stellar population not earlier than
$\sim$1--2~Gyr ago. Several similar type  very gas-rich `unevolved'
dwarfs were already found among the Lynx-Cancer void least massive galaxies
\citep{J0926,PaperIII,Triplet,U3672}.

For a deeper study of DDO~68, in this paper we analyse and combine several new
observations of this galaxy. They include the BTA long-slit spectroscopy and
photometry of the star-forming region Knot~3 containing the LBV, the BTA
FPI H$\alpha$ data on the ionized gas
kinematics in the `Northern ring' (supergiant shell) of DDO~68 in conjunction
with the {\it HST} H$\alpha$ images of giant shells. To follow the LBV variability,
we also use the photometry of Knot~3 in  archive {\it HST} and SDSS images.

Evolution and death of  extremely metal-poor massive stars are
among the principal issues for modelling galaxy formation and evolution
during the first 1~Gyr  after the Big Bang (e.g., \citet{Barkana01}).
While the state-of-art stellar evolution models including those with 
fast rotation have significantly advanced over the past decade (e.g.,
\citet{Szecsi15,PARSEC} and references therein), there is still no
direct comparison between
the model predictions and properties of true extremely metal-poor massive
stars. The main reason for this is the lack of such stars
in the local Universe accessible for sufficiently detailed studies.
One needs to wait for the qualitative progress in  new instruments and
methods at the extremely large next generation of optical telescopes coming into
operation in the 2020s.
To pave the path for the next generation telescope studies of the lowest-metallicity
massive stars,  we present the list
of the DDO~68 most luminous stars related to  six record-low metallicity
regions of the current/recent SF.

The layout of the paper is as follows. In Sec.~\ref{sec:obs}, we present
all the observational data used and briefly describe their reduction.
 In
Sec.~\ref{sec:LBVphoto}, main results on the photometry of Knot~3 and the LBV
(DDO68-V1) and separation of the LBV light from that of the underlying \HII-region
are presented.
Sec.~\ref{sec:LBVspec} presents spectral data on the LBV and its
variations. In Sec.~\ref{sec:shells}, main results on H$\alpha$ shells and
the Giant Super Shell are summarized.
Sec.~\ref{sec:liststars} presents the list of the most luminous stars
within the Northern Ring situated in the regions of extremely low gas
metallicity.
Sec.~\ref{sec:dis} is devoted to the discussion of new results,
their comparison with the previously available data and understanding them in a wider
context.
In Sec.~\ref{sec:summ} we summarize the new results and draw
the main conclusions of this DDO~68 study.
The distance to DDO~68 is adopted according to the TRGB-based
estimate from Makarov et al. (2016, in press) of 12.75~kpc (or $\mu$ =
30.53~mag). The
latter distance is very close to those from \citet{Cannon14,Sacchi16}.
The respective  scale is 62~pc in 1 arcsec.

\section[]{OBSERVATIONAL DATA AND REDUCTION}
\label{sec:obs}

\subsection{{\it HST} data}

DDO\,68 was observed with the {\it HST} using the Advanced Camera for Surveys
($ACS$) (GO~11578, PI A.~Aloisi). Deep exposures were acquired  with the filters
$F606W$ (7644~s, 2010.05.02) and $F814W$ (7644~s, 2010.04.27).
A deep image of the galaxy was also obtained in the F658N narrow-band filter,
centred on the H$\alpha$ line, for a total integration time of 2388 s on May 2, 2010.
For brevity, hereinafter we call
these three bands as $'V'$, $'I'$, and H$\alpha$.

We use the $ACS$ module of the {\sc DOLPHOT} software package by
A.~Dolphin\footnote{http://americano.dolphinsim.com/dolphot} to obtain the
photometry of the resolved stars
as well as to run artificial star tests to characterize the completeness and
uncertainty in the measurements. The data quality images were used to mask
bad pixels. The stars with high-quality photometry only were used in the
analysis. We have selected the stars with a signal-to-noise (S/N) ratio of at least
five in both  broad filters and $\vert sharp \vert \le 0.3$. The detailed
analysis of data, photometry, and resulting colour magnitude diagram are
given in our previous paper (Makarov et al., 2016, MNRAS, in press).

We had already noticed in PKP05 that  star-forming regions along the
`Northern ring' visible well in the BTA H$\alpha$ image can actually be
a manifestation of the induced SF in front of a supershell.
The most detailed H$\alpha$ images obtained at the {\it HST} uncover  fine
arc-like and shell structures in the `Northern ring' SF regions. This provides an 
opportunity for a better understanding of  ionized gas shells and their relation
to  youngest massive stars including the DDO~68 LBV.

\subsection{BTA data}

Spectral and photometric data at the BTA (SAO RAS 6-m telescope) were obtained at the
prime focus with the multimode device SCORPIO \citep{SCORPIO}. The parameters
of observations are shown in the Journal  (Table~\ref{Tab1_BTA}) and, for dates
prior to January 2009, they are presented in our papers PKP05;
\citet{LBV} in more detail.
For the fresh spectrum obtained in January 2009,
the slit width was of 1.0 arcsec. During the night, we observed the
spectrophotometric standard Feige~34. Spectral resolution for the grisms
VPHG550G and VPHG1200G was FWHM $\sim$12~\AA\ and
$\sim$5.5~\AA\ respectively. The photometry of the DDO~68 \HII\ region Knot~3 in 2009, 2015,
and 2016 as well as in January 2005 is based on the local standards --
several sufficiently bright stars in the SDSS images of DDO~68.
Their $g,r,i$ magnitudes were transformed to the Johnson-Cousins $B,V,R$
magnitudes according to the relations by \citet{Lupton05}.

Spectral and photometric observations of DDO~68 Knot~3 in November 2004
and January 2005, when the LBV features were below the detection level,
allowed us to define the level of the underlying \HII\ region light. The
latter is the basic parameter in disentangling the fraction of the LBV light,
when we register the integrated light
of both the \HII\ region and the LBV with  ground-based telescopes.

We follow the standard processing of 2D spectra, which include debiasing,
flat-fielding, wavelength and flux calibration, similar to that
described in detail in PKP05.
From the prepared 2D spectra, we extracted the fragment with a full
length of $\sim$5~arcsec
which included all the visible light of the \HII\ region Knot~3 falling within the
long slit. We convolved these spectra with the $B$ and/or $V$ passbands to use
them in the analysis of the Knot~3 variability. Appropriate corrections for
the lost part of light were performed according
to the procedures described in Sec.~\ref{ssec:BTAphoto}.

The  observations of DDO~68 ionized gas kinematics
in the H$\alpha$ emission line were conducted
with the same SCORPIO device in the scanning FPI mode as described in
\citet*{Moiseev10} and \citet{Moiseev2014}. The reduced data were combined
into the data cube, where each 0\farcs7 pixel in the 6\farcm1
field of view
contains a 36-channel spectrum sampled with 0.37~\AA\ (17~\kms) per channel.
Spectral resolution was $FWHM= 42$~\kms\, which corresponds to the velocity
dispersion $\sigma=18$~\kms. \citet{Moiseev2014} presented the study of the
large-scale disc rotation based on the velocity field derived from the FPI
data;  \citet*{Moiseev2015} have published the maps of the ionized
gas velocity dispersion. Here we consider only the data set obtained with
the best weather conditions. The final angular resolution corresponds to a
seeing of
1\farcs8 in comparison to \citet{Moiseev2014} and \citet{Moiseev2015} with
 deeper data but with a worse 2\farcs7 resolution.

\begin{table}
\begin{center}
\caption{Log of  observations of DDO~68 Knot~3 at the BTA}
\label{Tab1_BTA}
\hoffset=-2cm
\begin{tabular}{l|l|l|l|c|c} \hline  \hline \\ [-0.2cm]
\MC{1}{c|}{Date} &
\MC{1}{c|}{Grism}&
\MC{1}{c|}{Expos.}&
\MC{1}{c|}{PA} &
\MC{1}{c|}{$\beta$\arcsec}&
\MC{1}{c}{Air}  \\

\MC{1}{c|}{ } &
\MC{1}{c|}{or band} &
\MC{1}{c|}{time, s}&
\MC{1}{c|}{ } &
\MC{1}{c|}{} &
\MC{1}{c}{mass}\\

\MC{1}{c|}{(1)} &
\MC{1}{c|}{(2)} &
\MC{1}{c|}{(3)} &
\MC{1}{c|}{(4)} &
\MC{1}{c|}{(5)} &
\MC{1}{c|}{(6)} \\
\\[-0.2cm] \hline \\[-0.2cm]
 2004.11.09  & VPHG550G & 1$\times$900 &-57 & 0.8 & 1.08 \\ %
 2005.01.12  & $V$      & 3$\times$600 &    & 1.7 & 1.07 \\ 
 2005.01.12  & $R$      & 3$\times$600 &    & 1.9 & 1.20 \\ %
 2005.01.13  & VPHG550G & 3$\times$900 &-57 & 1.2 & 1.09 \\ 
 2006.12.30&FPI H$\alpha$&36$\times$120&    & 1.8 & 1.04 \\ 
 2008.01.11  & VPHG550G & 5$\times$900 &-26 & 1.2 & 1.20 \\ 
 2008.02.04  & VPHG1200G& 5$\times$900 &-26 & 1.3 & 1.13 \\ 
 2009.01.21  & $V$      & 1$\times$300 &    & 1.3 & 1.19 \\ 
 2009.01.21  & VPHG1200G& 6$\times$1200&-26 & 1.3 & 1.12 \\ 
 2015.01.14  & $B$      & 2$\times$150 &    & 2.1 & 1.05 \\ 
 2015.01.14  & $V$      & 2$\times$180 &    & 2.1 & 1.05 \\ 
 2015.01.14  & $R$      & 2$\times$180 &    & 2.1 & 1.05 \\ 
 2016.01.15  & $B$      & 5$\times$180 &    & 2.2 & 1.10 \\ 
 2016.01.15  & $V$      & 5$\times$180 &    & 2.2 & 1.14 \\ 
 2016.01.15  & $R$      & 5$\times$180 &    & 2.2 & 1.07 \\ 
\hline \hline \\[-0.2cm]
\end{tabular}
\end{center}
\end{table}

\subsection{SDSS data}

SDSS images of DDO 68 in the filters $u,g,r,i,z$ were obtained at the epoch
2004.04.16. Since the images presented in the SDSS database are fully reduced,
 we only need to measure  the
underlying background before we perform photometry in round apertures.
This requires special care since the object of interest (Knot~3 with the LBV)
is situated near  other regions of recent and current star formation.
Therefore, we measure the background with small apertures in many adjacent
regions, which exclude any contribution of  neighbouring diffuse or stellar
objects.

\section[]{LIGHTCURVE OF THE LBV}
\label{sec:LBVphoto}

\subsection{Determination of the \HII\ knot contribution from the  {\it HST} data}
\label{HST-data}

High-angular-resolution {\it HST} images of this region give us a unique
opportunity to disentangle the fluxes of the underlying \HII\ region and
the star-like object, to which the major light contribution comes from the
LBV. This allows, on the one hand, to get the direct estimate of the LBV
luminosity at that epoch and, on the other hand, gives us the integrated
$V$ and $I$ magnitudes of the star-forming region Knot~3 without the LBV
contribution. They are assumed to correspond to the minima of the Knot~3
lightcurve.

Since the LBV (together with the expected young star cluster, to which it
belongs) is a star-like in the ACS images, we have measured its individual
total magnitudes in the {\it HST} $F606W$ and $F814W$  filters and
transformed them to the Johnson-Cousins $V$ and $I$ bands according to
the relations suggested in \citet{Sirianni05}. In the
{\it HST} images, we also performed  the aperture photometry similar to our ground-based
photometry, that is integrated photometry for the LBV together with all the
light of the underlying SF region Knot~3 (the sum of compact ionizing star cluster
and the extended \HII\ region along with many fainter star-like objects
within this region, see Fig.~\ref{fig:knot3}).
For the latter estimate, the radius of the round aperture centred on the LBV
was taken equal to 2\farcs5, very close to that of our photometry of Knot~3
in the BTA and SDSS images. The background was adopted as the average of the
light in many small apertures (to avoid the background stars) around the
measured region.

In the standard $BVRI$ system, we obtain the following
magnitudes: $V_{\rm LBV} = $20.05$\pm$0.08, $I_{\rm LBV} =$ 19.93$\pm$0.03 for the LBV itself.
The Milky Way extinction is not applied to the apparent magnitudes across the
paper including those in Table~\ref{t:lum.list}.
For the integrated magnitudes of this region, we obtain:
$V_{\rm LBV+Kn.3} =$ 19.26$\pm$0.08 and
$I_{\rm LBV+Kn.3} =$ 19.16$\pm$0.03. The rms uncertainties, 0.08$^m$ and
0.03$^m$, come from the related errors in the transformation formulae by
\citet{Sirianni05}.
The respective absolute magnitude
of the LBV at this epoch was $M_{\rm V,LBV} =$ --10.53.

\subsection{BTA photometry and algorithm to recover the intrinsic
LBV magnitudes}
\label{ssec:BTAphoto}

We performed the photometry of DDO~68 Knot~3 containing the LBV in the images
obtained in the BTA observations on 2005.01.12 ($V$, $R$),
2009.01.21 ($V$) and 2015.01.14, and 2016.01.15 ($B$, $V$, and $R$). The primary
processing
of images was carried out with the standard pipeline with the use of both  IRAF and MIDAS.
It included the removal of cosmic ray hits, bias, and flat correction. The background
was determined as an average for several small regions around Knot~3 within
a distance of 5~arcsec and then was subtracted from the image before
 measuring the Knot~3 instrumental magnitudes. The last were transformed
to the $B,V,R$ magnitudes via  average zero-points obtained on four local
standards -- stars with $V$ in the range of 17.5$^m$ to 18.8$^m$ selected in
this region based on the SDSS images. Their magnitudes were measured in
the $g,r,i$ filters and transformed to $B,V,R$ with the corresponding formulae from
\citet{Lupton05}. The related uncertainties in the average zero-points of
these filters are  0.014$^m$, 0.020$^m$, and 0.024$^m$. In each case, to
obtain the total magnitude of Knot~3, we used the round aperture with
$r =$2\farcs5.

To extend the photometric time series, we also used the convolution of the BTA
extracted 1D spectra of Knot~3 with the respective passbands of  the $B$ and $V$
filters for spectra with the grism VPHG550G (the range 3600--7000~\AA) and
$B$ filter for the grism VPHG1200G (the range 3600--5700~\AA).
Due to a rather small width of the long slit (1~arcsec), the substantial part
of the light was lost. Therefore, to obtain the total Knot~3 magnitudes from
those acquired with the spectra convolution of its part, one needs to make
correction for the light loss. This requires an adequate model of the complex
light distribution and additional calculations.

We calculated all necessary corrections for each of the used spectrum
convolutions taking the value of seeing ($\beta$) as the input parameter during
the respective observation. For this we used the original {\it HST} image
of Knot~3 with a FWHM of $\sim$0\farcs1 and performed its Gaussian smoothing
to reproduce seeings for the respective spectra for the $\beta$ range from
0\farcs8 to 1\farcs8.
For each of such smoothed images, we overlaid the rectangle of the slit
of 1\arcsec $\times$ 5\arcsec. Separately for the underlying \HII\ region and
the star-like LBV, we integrated their light within the slit and derived the
lost light fraction with respect to the total light.
The derived correction factors (i.e., multiplicators to estimate the full
flux from that appeared in the slit) vary from 1.36 to 2.07 for $\beta$
between 0\farcs8 and 1\farcs8 for a star-like object, and respectively, from
2.33 to 2.85 for this concrete underlying \HII\ region.
To estimate corrections of the Knot~3 magnitudes derived via the convolution
of spectra from \citet{IT09}, we used their respective slit size of
1\farcs5 by 3\farcs5 for $\beta =$ 0\farcs8$\pm$0\farcs2.

In the following algorithm, we use the `known' total $V$ magnitude of the
underlying \HII\ region as estimated from the deconvolution of the {\it HST} image
($V$(\HII)=20.12$^m$). This number was also supported by the independent BTA
photometry of Knot~3 on 2005.01.12 at the epoch, when no tracers of the LBV
light were seen in its spectrum (obtained in the next night).
From the spectra convolution with the respective filter passbands ($V,B$),
using the algorithm below, we recover (at a first approximation) the total
magnitudes of Knot~3 in $B,V$ and separately the $V$ magnitude for the LBV.

The algorithm includes the following steps. We start, for example, from the
$V$ magnitude of the convolved spectrum obtained on 2008.01.11 under
$\beta$=1\farcs2, for which we derive $V$(conv)=20.83$\pm$0.02. The
respective factors of loss on the slit are 1.53 for a star-like object and
2.5 for the Knot~3 \HII\ region. From the known magnitude $V$(\HII)=20.12$^m$,
applying the respective correction, we estimate its contribution to $V$(conv)
on the level of $V$(\HII)(conv) = 21.12$^m$.  The difference in light between
$V$(conv)=20.83$^m$ and $V$(\HII)(conv) = 21.12$^m$ is assigned to the light
of the LBV within the slit. From simple calculations, it corresponds to
$V$(LBV)(conv) = 22.40$^m$. Then, after the appropriate correction for the
slit light loss, we derive the true $V$(LBV)=21.94$^m$. Finally, summing up
the fluxes for $V$(\HII)=20.12 and that for the LBV, we derive the integrated
light for Knot~3: $V$(Kn.3)=19.93$^m$.

Different calculations are applied, when we directly measure the total light
of Knot~3 via the aperture photometry and estimate the light of the LBV.
We again use the basic parameter $V$(\HII)=20.12$^m$ and assign all extra light
to that of the LBV.

\subsection[]{SDSS results}
\label{ssec:res_SDSS}

The results of the SDSS total magnitude measurements of Knot~3 in a round
aperture with $r = $2\farcs5 on 2004.04.16 are as follows:
$g$ = 19.85$\pm$0.02,   $r$ = 19.83$\pm$0.02. According to the transformation
formula from \citet{Lupton05}, they translate into $B$ = 20.08$\pm$0.03,
$V$ = 19.84$\pm$0.03, and $R$ = 19.73$\pm$0.03.

\begin{figure}
  \centering
 \includegraphics[angle=-90,width=7.5cm, clip=]{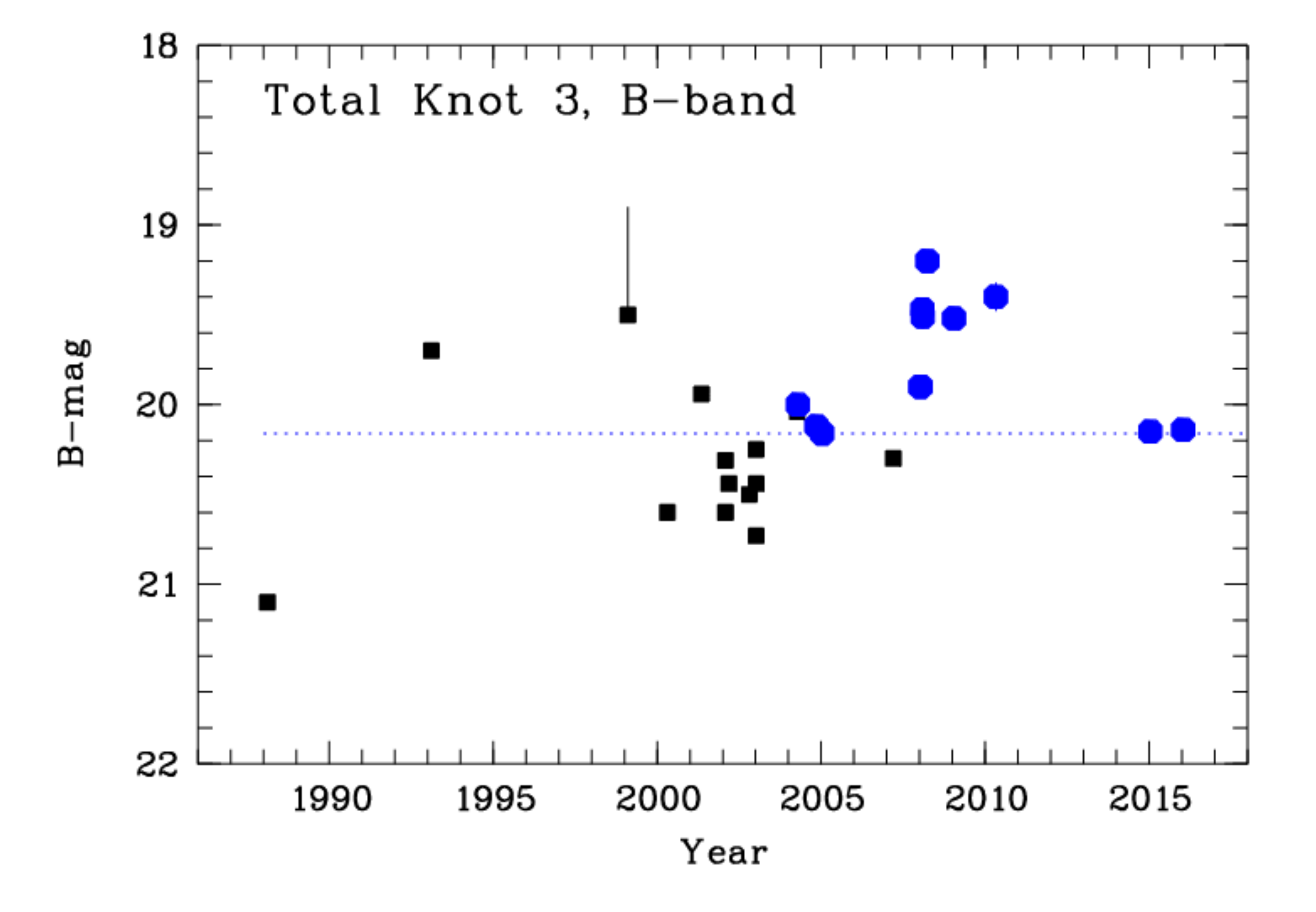}
 \includegraphics[angle=-90,width=7.5cm, clip=]{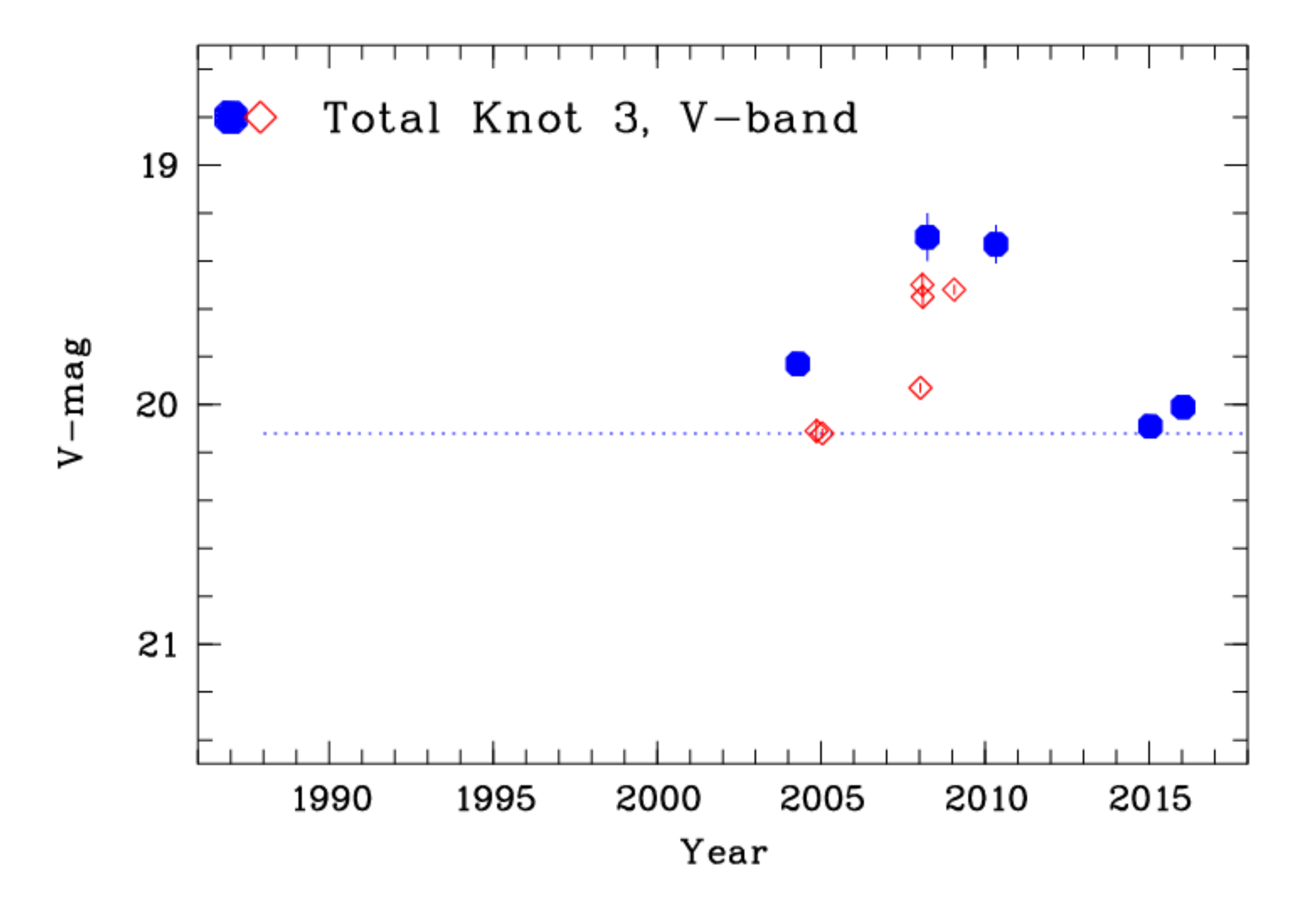}
 \includegraphics[angle=-90,width=7.5cm, clip=]{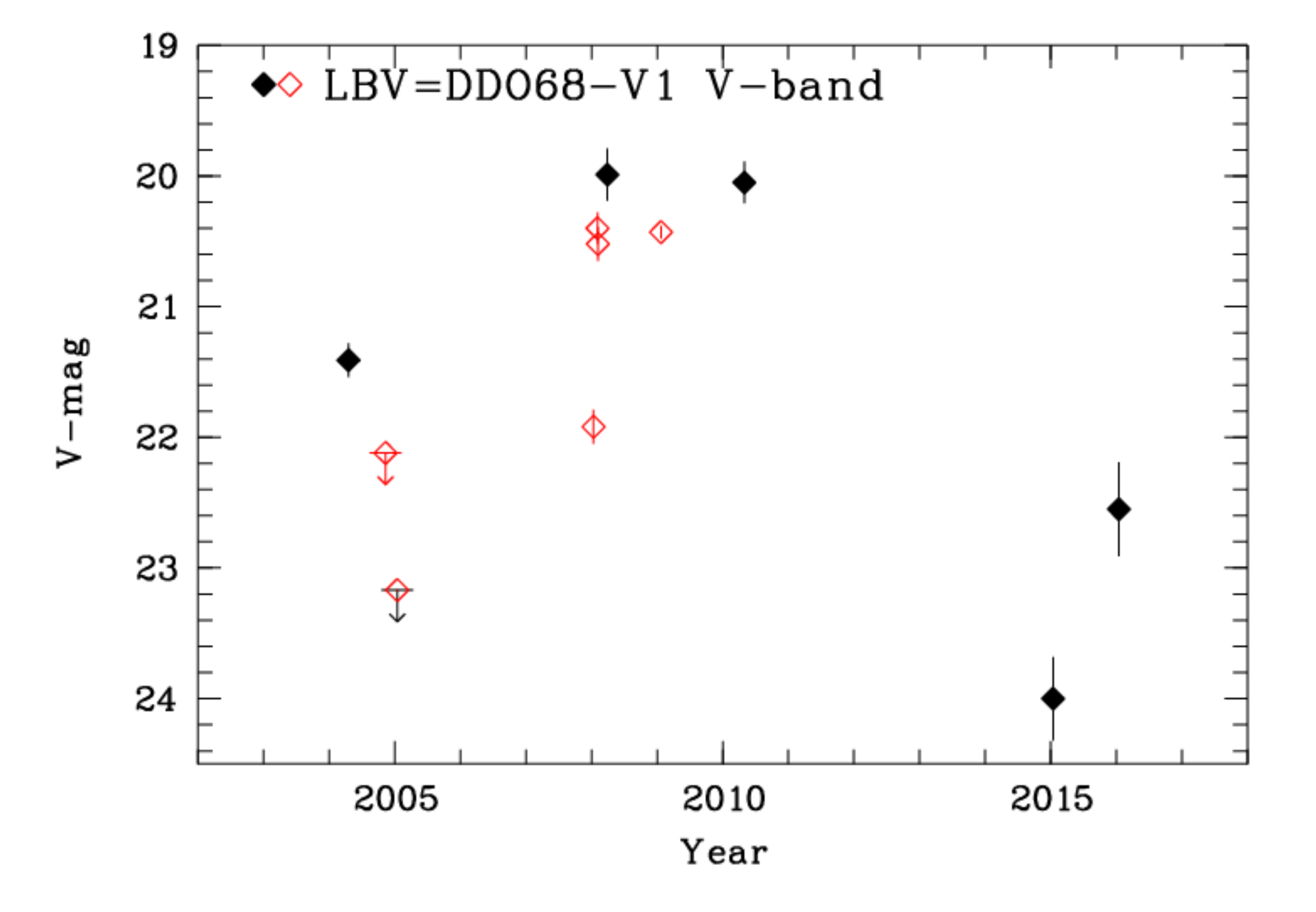}
  \caption{{\bf Top panel.} The lightcurve of SF Knot~3 in DDO~68 in the $B$
  band since 1988. The filled (black) squares are from
   \citet{Bomans11,BW11}. The filled (blue) hexagons are  new data.
   They all are for an aperture with $r$ = 2\farcs5 (see Table~\ref{tab:photo} and
   its description). The part of $B$-magnitudes is based on the $V$ band data and
   $B-V$ colour obtained  via the  convolution of the object spectra.
  A dotted line at $B$ = 20.16 corresponds to the light of Knot~3, when the LBV
   is not seen in the spectrum.
   {\bf Middle panel.} The same but for the $V$-band. The data obtained from the
  {\it HST} and SDSS images are transformed to $V$ (see the text). The rest of
   the data is from our direct BTA photometry (filled lozenges), supplemented by
   points based on the BTA, MMT, and APO spectra convolution corrected for the light
   loss on slit  as described in the text (open red lozenges).
   The dotted line at $V$ = 20.12 corresponds to the light of the entire
   Knot~3 when the LBV contribution in the Knot~3 spectrum is not visible.
{\bf Bottom panel.} The light curve of the LBV in the $V$-band (filled lozenges).
  With the exception of one direct photometry in the {\it HST} image, all
  other magnitudes are derived as the `residual light' via subtraction of
 the constant luminosity of the underlying \HII\ region ($V$=20.12) from the
 previous lightcurve. Lozenges with arrows indicate 3$\sigma$ upper limits.
 Points obtained from the convolution of Knot~3 spectra are shown with open
  red lozenges.
}
	\label{fig:lightcurve}
 \end{figure}

\subsection{DDO~68 LBV light variability}
\label{ssec:variable}

Turning to DDO68 LBV (also DDO68-V1), it is worth remembering the other most
studied low-metallicity extragalactic LBVs considered to be caught in the
phase of giant eruption. One of them is NGC~2363-V1 (in the \HII\ region with
12+$\log$(O/H)=7.9)
\citep{Drissen97,Drissen01,Petit06}. The other one is the LBV in PHL293B
residing in the \HII\ region with 12+$\log$(O/H)=7.71 \citep{IT09,Izotov11}.
The former object was well observed with the {\it HST} during the period of
its maximal light when it brightened by $\sim$3.5~mag and had reached
$M_{\rm V} \sim$--10.5. The latter LBV was found in the very bright phase,
with $M_{\rm V} \sim$--12, close to the maximum known for LBV luminosities.

In the context of the LBV study, their giant eruptions with the maximum mass-loss
rates are especially interesting. It is tempting to check whether DDO68-V1
approached this phase during the studied time interval.
The whole dataset on the DDO~68 LBV variability is
very limited. With the only exception of the {\it HST} images, all other
measurements deal with the integrated light of the star-like LBV itself and
the extended light of the underlying star-forming region Knot~3 obtained at
typical seeings of 1\arcsec -- 2\arcsec. Fortunately, as shown in
Section \ref{ssec:BTAphoto},  we have its emission as that of a separate
object well measured in the {\it HST} images (Subsection \ref{HST-data}).
Therefore, one can disentangle the total measured fluxes into two components
and derive the
contribution of the LBV light with the acceptable accuracy.

Very few magnitude estimates of Knot~3 in DDO~68 have been known during about the last
60 years.  Most of them are presented by \citet{Bomans11,BW11} in the
$B$ and $R$ Johnson-Cousins bands. These estimates encompass the period
from 1955 [two measurements from the Digitized Sky Survey (DSS) - digitized
`Palomar Observatory Sky Survey One' -- POSS1 photoplates] to
2007.  During the period of 1988--2007, most of the data were obtained via
CCD measurements.
They also include the flux estimates derived from the `convolution' of the
spectra by \citet{IT09} in February and March 2008.
The maximum light of the Knot~3 region, as shown on \citet{BW11}
lightcurve, was as high as $B \sim$18.9 (February 1999). However,  the `error-bar' of this point drawn in their Figure~2 (with no comment on its
meaning) assumes that in fact it can be as faint as $B \sim$ 19.5.

Due to the absence of any information on seeings, used apertures and the
background subtraction, this dataset is difficult to use for examination
of the range of light variations of the LBV itself.
Their total range of the observed luminosity variations of Knot~3  is
$\delta B$(Kn.3) $\sim$2$^{m}$, if one adopts the brightest state of
$B \sim$19.1 (POSS1 photoplate magnitude from 1955) and the lowest state with
$B \sim$21.1 in 1988. Both these extreme points are very important in
further analysis. However, their reliability, especially for the minimum,
can be questioned in view of our measurement of the minimal light of Knot~3
at the level of $V =$ 20.12$^m$ (see below).


\begin{table*}
\caption{Summary of our $B,V,g$ total magnitude estimates
for DDO~68 SF region~3 and the LBV
\label{tab:photo}
}
\begin{tabular}{lllllllrll}
\hline
\multicolumn{1}{l}{Date}  &
\multicolumn{1}{c}{$B$} &
\multicolumn{1}{c}{$\sigma_{\rm B}$} &
\multicolumn{1}{c}{$g$} &
\multicolumn{1}{c}{$\sigma_{\rm g}$} &
\multicolumn{1}{c}{$V$} &
\multicolumn{1}{c}{$\sigma_{\rm V}$} &
\multicolumn{1}{c}{$V$(LBV)} &
\multicolumn{1}{c}{$\sigma_{\rm V}$(LBV)} &
\multicolumn{1}{l}{Ref. Notes}     \\
 \hline \\[-0.2cm]
040416  &20.00  &0.05  &19.85   & 0.02  & 19.83 & 0.03 &21.41      & 0.13  & SDSS phot \\
041109  &20.12  &0.06  &20.08   & 0.04  & 20.11 & 0.05 &$\geq$22.12& 3$\sigma$ up.lim. & BTAcv    \\  
050113  &(20.16)&0.05  &20.11   & 0.05  & 20.12 & 0.02 &$\geq$23.23& 3$\sigma$ up.lim. & BTA phot+cv\\
080111  &19.90  &0.03  & 19.91  & 0.03  & 19.93 & 0.02 & 21.92     & 0.13  & BTA cv   \\
080202  &(19.47)&      &        &       & 19.50 & 0.05 & 20.40     & 0.12  & APO IT09 cv   \\
080204  & 19.51 &0.07  &19.55   & 0.05  & 19.55 & 0.05 & 20.52     & 0.13  & BTA cv   \\ 
080328  &(19.20)&0.10  &        &       & 19.30 & 0.10 &19.99      & 0.20  & MMT IT09 cv \\
090121  & 19.52 &0.02  &(19.53) &       & 19.51 & 0.01 &20.43      & 0.03  & BTA phot+cv \\
100502  &(19.40)&0.08  &        &       & 19.33 & 0.08 &20.05      & 0.08  & HST phot \\
150114  & 20.25 &0.01  &(20.09) &       & 20.09 & 0.01 &24.00      & 0.32  & BTA phot \\
160115  & 20.14 &0.02  &        &       & 20.01 & 0.03 &22.55      & 0.32  & BTA phot \\ 
\\[-0.25cm] \hline \\[-0.2cm]
\multicolumn{10}{l}{cv -- estimates via spectrum convolution with the filter passband and the respective on-slit loss correction.} \\
\multicolumn{10}{l}{(): expected values based on the other bands and the measured colours; IT09 - Izotov \& Thuan, 2009. } \\
\end{tabular}
\end{table*}

We perform a more detailed analysis of the available photometric
estimates of Knot~3 using our photometry at the BTA, the SDSS image database,
and {\it HST}/ACS images in the $F606W$ and $F814W$ filters.
For the subsequent analysis of the lightcurve, we aim to have the {\it HST}
measurements to be consistent within the adopted uncertainties with the BTA
measurement of
$V_{\rm LBV+Kn.3}$=20.12$\pm$0.02 on 2005.01.12. At this epoch, the LBV was
too faint and did not show up in the spectrum of Knot~3. Hence, the
$V$ and $R$ magnitudes measured  at this epoch should be treated as a good
approximation for that of the underlying \HII\ region. To make the {\it HST}
and BTA values agree, we adopt the integrated {\it HST} magnitude
$V_{\rm LBV+Kn.3} =$ 19.33.
The latter value is within 1$\sigma$ uncertainty from the nominal value
19.26$\pm$0.08.

To constrain the light of the LBV in the nights of January 12-13, 2005, we
used the following approach. Taking the cited uncertainty of the aperture
photometry $\sigma_{\rm V} =$ 0.02$^m$, we adopt that the light contribution
of the LBV at that epoch with  high confidence does not exceed the
3$\sigma$ level, that is $\sim$6~\% of the `basic' $V$-band luminosity of
Knot~3, corresponding to $V$=20.12$^m$. Hence, the related upper limit
corresponds to $V_{\rm LBV} \geq$ 23.23$^m$.

In this `faint' state, in the {\it HST} image in Fig.~\ref{fig:knot3},
the LBV would look
like  star no.~9 in its nearest neighbourhood, or fainter. The brightest
directly observed state of the LBV, as derived from the {\it HST} image
photometry, corresponds to $V$(LBV) = 20.05$^m$. This implies that during
the period of about 5 years (2005-2010, see Table~\ref{tab:photo}) the
LBV has brightened by more than 3.1$^m$.

Our BTA measurement of the total light of Knot~3 and the LBV
on 2015.01.14  ($V_{\rm total}$ = 20.09$\pm$0.01)
corresponds to the luminosity which is 3$\sigma$ above its `basic'
level.
The derived contribution of the LBV  corresponds to
 $V_{\rm LBV} =$ 24.0$\pm$0.32.
This implies that between May 2010 and January 2015 the LBV $V$-band
luminosity decreased by a factor of $\sim$40, or faded by
$\delta V\sim$4.0$\pm$0.32$^m$. Accounting for its rather large uncertainty,
the more robust estimate is  $\delta V$ $\gtrsim$ 3.7$^m$.

For better understanding of  the character of the lightcurve of Knot~3 and the LBV
during the last decade, we have analysed all available data including own
imaging with the BTA, images from the SDSS and {\it HST} databases and the BTA
spectra convolution.
We also convolved the spectra from \citet{IT09} obtained on February 2 and
March 28, 2008, when the LBV was in its bright phase. We employed the approach
described and illustrated above at the end of Section~\ref{ssec:BTAphoto}.
No information is available on the seeing ($\beta$) for
observations of the Knot~3 spectra
by  \citet{IT09}. We adopted the median value of
$\beta =$0\farcs8\ typical of the Apache Point Observatory (APO) 3.5-m and
MMT telescopes.
Its r.m.s. $\sigma_{\beta} \sim$ 0\farcs2 results in a relatively small
additional scatter
in the derived magnitudes of Knot~3. This is included in our final errors.
All the derived
estimates are presented in Table~\ref{tab:photo} along with  other data
and shown in Fig.~\ref{fig:lightcurve}.

As one can see, the LBV approached its maximum observed brightness already
at the beginning of 2008. During  next two years, there were only two
additional points, in January 2009 (BTA) and in May  2010 ({\it HST}),
which suggested that its light
varied close to that maximum level corresponding to $M_{\rm V}$(LBV)
of about --10.5$^m$. It appears that there are no published images of
Knot~3 and/or its light measurement between May 2010 and January 2015.

Table~\ref{tab:photo} summarizes all our photometric data on the
integrated magnitudes of Knot~3 and those related to the LBV.
In column~1, the date of observation is given. In columns 2 and 3 -- the total
$B$ magnitude and its adopted uncertainty $\sigma_{\rm B}$. The part of the data
in parentheses are the estimates recalculated from the $V$ magnitudes in the same
night, either directly from the $B-V$ of convolution or indirectly via
the typical object colours for the similar brightness. In Columns 4 and 5,
the similar values are presented for $g$ filter. Columns 6 and 7 show
the adopted integrated $V$-magnitudes and $\sigma_{\rm V}$. Columns 8 and 9
give the LBV $V$ magnitudes and their errors.
In Column 10, we give info on the telescope and the origin of parameters in
the previous columns.

The amplitude of the DDO~68 LBV variability, found in the course of this study, 
$M_{\rm V}$  from $\gtrsim$--6.95 to --10.58, is very large. This indicates
that during the period of roughly two to five-six years (certainly during the
years 2008-2010 and probably longer) it experienced a giant eruption.
If the maxima in the \citet{Bomans11} $B$-band lightcurve of Knot~3 in 1955
($B \sim$19.1) and 1999 ($B \sim$18.9, see Fig.~\ref{fig:lightcurve}) are
realistic, then such events could have repeated two or three times over the last 60
years.

By the way, it is worth noting that some of \citet{Bomans11} data seem
to contradict our $V$ band lightcurve. Indeed, the faintest total
magnitude of the SF region Knot~3 $V_{\rm tot}$ = 20.12 corresponds to the
lowest LBV luminosities when
it was below the detection level. Since the $B-V$ colour for the \HII\ region is
rather blue (say, $\lesssim$0.25~mag), one does not expect its $B$ magnitudes
to be fainter than $\sim$20.4. In fact, in the \citet{Bomans11} $B$-band
lightcurve, several such points are drawn, including the absolute minimum
point in 1988 at $B \sim$ 21.1.
Since the authors do not present details of their photometry, one could think
that the faintest values of their lightcurve are related to either
systematical difference due to their photometry in smaller than our
apertures or to some other unknown reasons, which only the authors can
determine.

\section{LBV spectrum}
\label{sec:LBVspec}

\begin{figure*}
  \centering
\includegraphics[angle=-90,width=7.5cm, clip=]{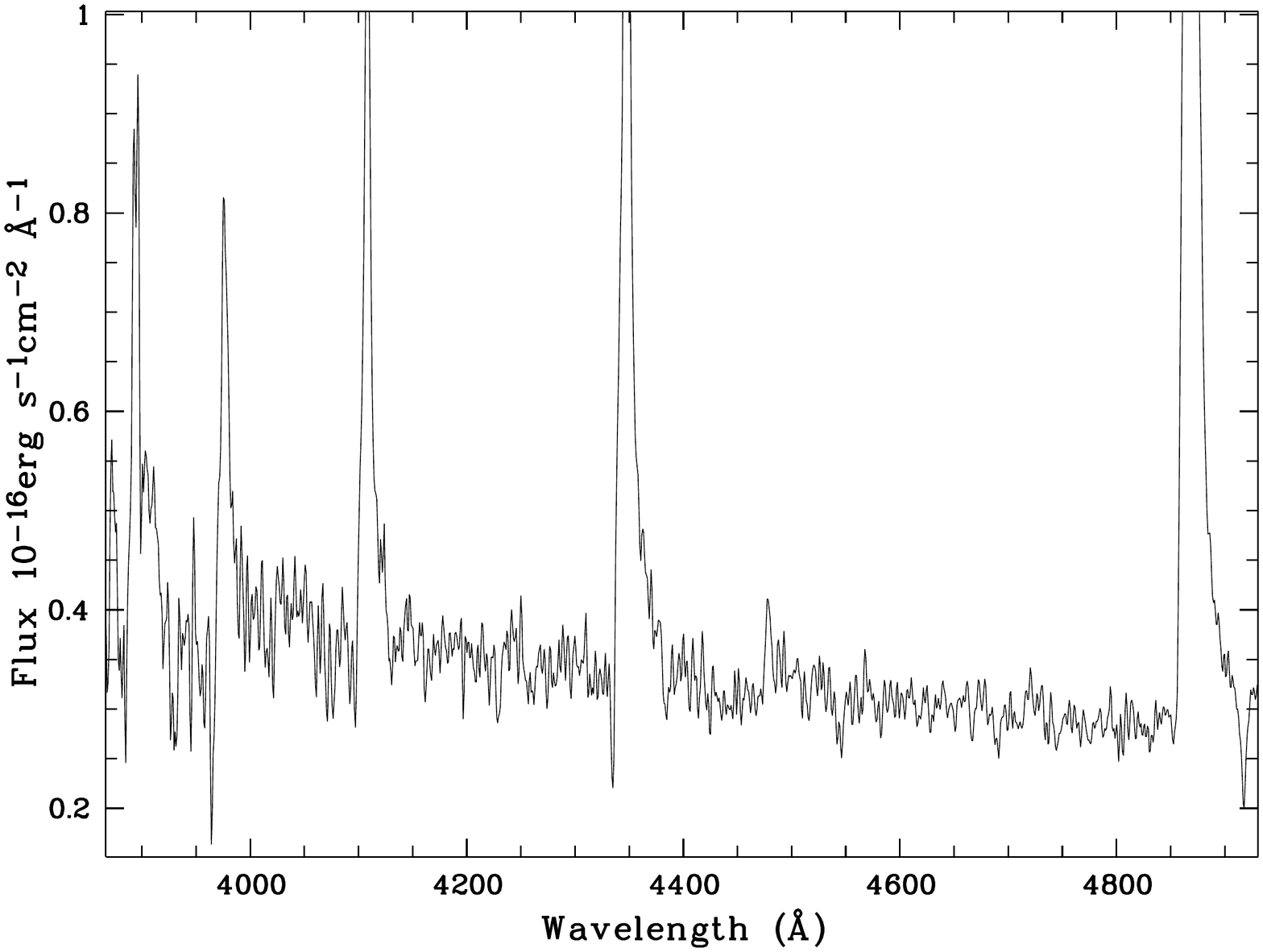}
\includegraphics[angle=-90,width=7.5cm, clip=]{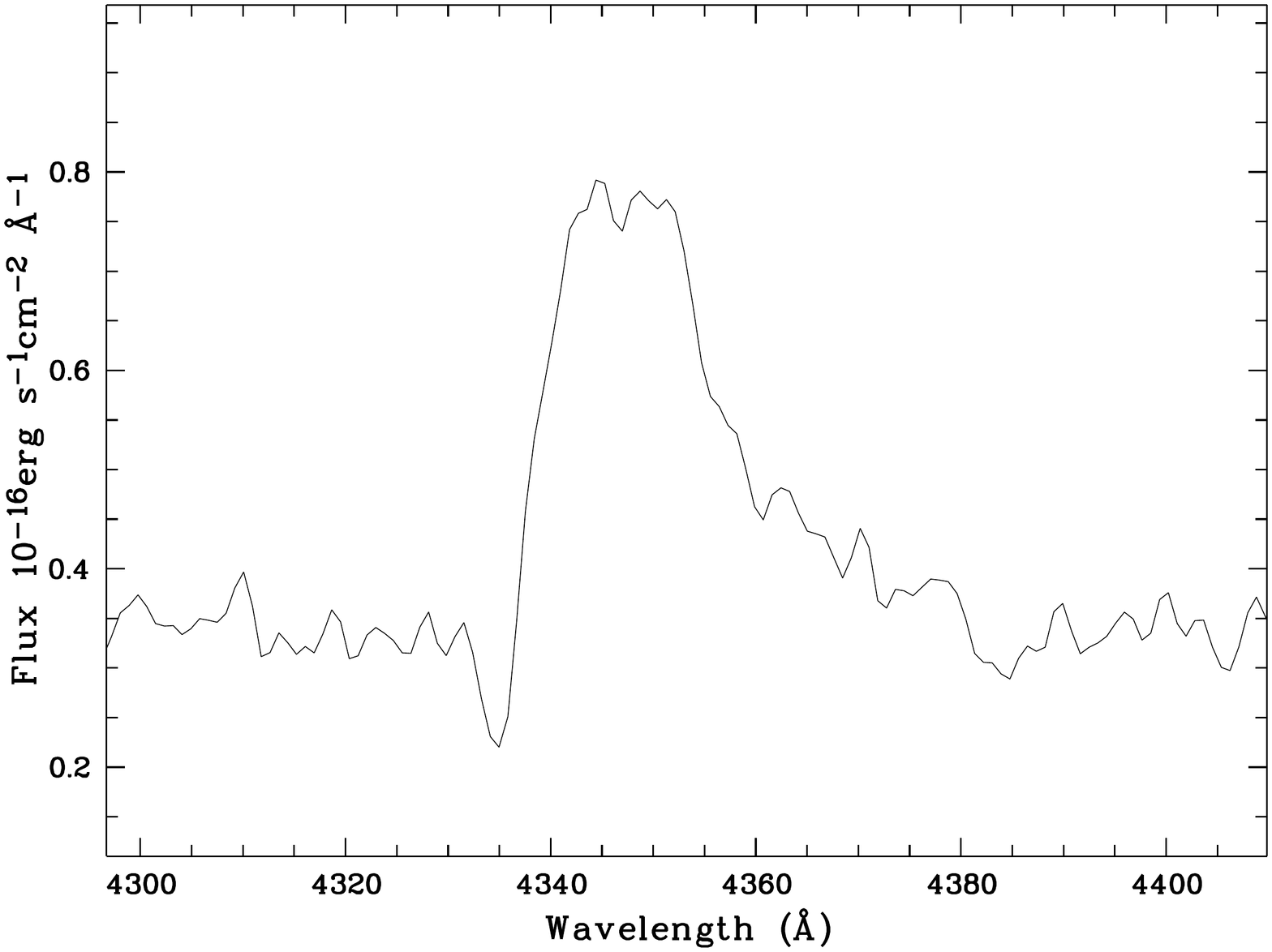}
  \caption{
{\bf Left panel}:
Zoomed in spectrum of \HII\ region Knot~3 in DDO~68 (including emission from
the LBV) obtained at BTA on Jan. 15, 2009. The features attributed to the LBV
wind/shell are well seen: broad (width $\sim$1000~\kms) low-contrast emission
line components in three Balmer lines and P-Cyg absorption components at
negative velocities ($\delta~V \sim$ 700--800~\kms).
{\bf Right panel}: The LBV H$\gamma$-line profile (after subtraction of the
'instrumental' profile of the narrow emission of the underlying \HII-region).
}
	\label{fig:spec3}
 \end{figure*}

Due to the volume limits, we do not show here the spectra of Knot~3
obtained in 2005 and 2008, which can be found in the discovery paper on
DDO~68 LBV \citep{LBV}. In Fig.~\ref{fig:spec3} (left panel), we show the
unpublished spectrum of Knot~3 obtained with the grism VPHG1200G on 2009.01.15.
Again, as in the BTA 2008 spectra, in difference to other known
LBVs only Hydrogen and Helium broad components of the LBV spectra are seen.
The only exception, similar in this respect to DDO~68 LBV, is
the LBV in PHL~293B with 12+$\log$(O/H) = 7.72 \citep{IT09}.
Also, the P-Cyg absorption in blue wings of Balmer lines is seen. The
latter is illustrated in the zoomed part of the object spectrum centred at
H$\gamma$ (Fig.~\ref{fig:spec3}, right panel).
Here the narrow 'instrumental' width component related to emission of
the underlying \HII\ region, was subtracted out. This was performed with
the standard MIDAS procedure 'blendfit' to fit Balmer lines with two Gaussian
profiles. The instrumental width (FWHM) of the narrow component
($\sim$5.5~\AA\ or $\sim$370~\kms\ at H$\gamma$) was adopted from the
measured linewidths of the strong symmetrical emission lines
[O{\sc iii}]$\lambda\lambda$4959,5007 attributed
to the underlying \HII\ region. This instrumental width is about three times
smaller than that of the broad lines related to the LBV.

For epoch 2009.01.15, the LBV
was in the high state, roughly 0.5$^m$ fainter than the firmly registered 
maximum for the total amplitude of more than 3.6~mag (see
Sect.~\ref{ssec:variable}). Therefore its line fluxes are suitable to combine
with other measurements to follow their behavior with the LBV variability.
Below, we perform their analysis,  
combining them with the
only additional published spectral data from \citet{IT09}.

\begin{table}
\begin{center}
\caption{Parameters of broad Balmer lines in the spectrum of 2009.01.21}
\label{tab:Balmer-param}
\hoffset=-2cm
\begin{tabular}{l|c|c|c|c} \hline  \hline \\ [-0.2cm]
\MC{1}{c|}{Name lines} &
\MC{1}{c|}{$\lambda_{\rm em}$} &
\MC{1}{c|}{$\lambda_{\rm abs}$}&
\MC{1}{c|}{FWHM} &
\MC{1}{c|}{$V-V_{\rm abs}$} \\
\MC{1}{c|}{ } &
\MC{1}{c|}{\AA } &
\MC{1}{c|}{\AA }&
\MC{1}{c|}{km/s} &
\MC{1}{c|}{km/s} \\
\MC{1}{c|}{(1)} &
\MC{1}{c|}{(2)} &
\MC{1}{c|}{(3)} &
\MC{1}{c|}{(4)} &
\MC{1}{c|}{(5)} \\
\\[-0.2cm] \hline \\[-0.2cm]
H$\beta$  & 4869.36 & 4856.31 &1079 & 806  \\
H$\gamma$ & 4347.48 & 4336.53 &1160 & 757  \\
H$\delta$ & 4108.20 & 4098.65 &1123 & 699  \\
\hline \hline \\[-0.2cm]
\end{tabular}
\end{center}
\end{table}


To date, there are four published spectra of \HII\ region Knot~3 during phases
when the LBV made the substantial contribution to its integrated spectrum.
They have all been obtained in the period from Jan.11 to Mar.28, 2008
(see Table~\ref{tab:lines-var}).
Only the LBV discovery spectrum on 2008.01.11 corresponds to its intermediate
phase, with $M_{\rm V,LBV} =$--8.6, when it was almost two magnitudes
fainter than in the maximum (see Table~\ref{tab:photo}). Three other spectra
are obtained at the LBV 'maximum' light, or at 0.4$^m$-0.5$^m$ fainter,
within $\sim$50 days prior to the maximum.
Despite the fact that it was acquired about 300 days after the maximum,
the new spectrum added here on 2009.01.21 corresponds to the LBV brightness of only
0.44$^m$ fainter than in the maximum. Therefore, apart from the very modest
amount of the LBV spectral data, they are also biased to the brighter quartile
of the LBV $V$ band light variations.

The only lines visible in these spectra are broad (FWHM $\sim$1000~\kms)
emissions of Hydrogen and Helium. No apparent emission lines of heavy
elements are detectable. Since we know of no other LBVs, except the aforementioned
next lowest metallicity LBV in PHL239B, whose spectra lacks the
heavy elements, we suggest that this is related to the very
low metallicity of its the main sequence progenitor. P-Cyg line profiles
clearly seen in the strong emission lines indicate the expanding wind/shell
with the terminal velocity of $\sim$800~\kms\ noticed earlier by \citet{IT09}.

Due to the very limited range of the LBV luminosity for spectral data and very
scarce time coverage, it is difficult to follow more or less gradual changes
of line fluxes. In such a situation one can only try to catch some trends
and to confront them with known similar objects. We present the LBV line
fluxes in Table~\ref{tab:lines-var} along with its adopted $V$ magnitudes.


\begin{table*}
\caption{Summary of the LBV (DDO68-V1) line variations on BTA, APO and MMT data
\label{tab:lines-var}
}
\begin{tabular}{lrrlrlrlrlrl}
\hline
\multicolumn{1}{l}{Date}  &
\multicolumn{1}{c}{$V_{\rm LBV}$} &
\multicolumn{1}{c}{$I$(HeI4471)} &
\multicolumn{1}{c}{$\sigma_{\mathrm 4471}$} &
\multicolumn{1}{c}{$I$(HeI5876)} &
\multicolumn{1}{c}{$\sigma_{\mathrm 5876}$} &
\multicolumn{1}{c}{$I$(H$\alpha$)} &
\multicolumn{1}{c}{$\sigma_{H\alpha}$} &
\multicolumn{1}{c}{$I$(H$\beta$)} &
\multicolumn{1}{c}{$\sigma_{H\beta}$} &
\multicolumn{1}{c}{$I$(H$\gamma$)} &
\multicolumn{1}{c}{$\sigma_{H\gamma}$} \\
 \hline \\[-0.2cm]
041109     &$>$22.12& $<$1.4 &        &$<$1.2&   0.5 &$<$1.0&0.5 &$<$0.5& 0.27&$<$0.5& 0.25 \\
050113     &$>$23.17& $<$0.5 &        &$<$0.6&   0.3 &$<$6.0&1.8 &$<$2.3& 0.7 &$<$1.3& 0.45 \\
080111     &  21.92 &    5.3 &     0.4&  10.2&   0.6 & 48.7 &9.0 & 22.2 & 4.6 & 19.2 & 4.6 \\
080202$*$  &  20.40 &        &        &      &       & 21.6 &0.5 & 10.9 & 0.6 &      &     \\
080204     &  20.52 &    0.7 &     0.2&   -- &       &      &    & 12.1 & 1.0 &  6.0 & 0.8 \\
080328$**$ &  19.99 &    0.4 &     0.2&      &       &      &    & 21.0 & 0.3 & 11.7 & 0.5 \\
090121     &  20.43 &    0.6 &     0.3&   1.3&   0.2 &      &    & 32.9 & 1.6 & 12.7 & 1.8 \\
\\[-0.25cm] \hline \\[-0.2cm]
\multicolumn{12}{l}{Line fluxes are in units 10$^{-16}$ erg cm$^{-2}$ s$^{-1}$. Upper limits at 2$\sigma$ level. Fluxes of Balmer emission lines} \\
\multicolumn{12}{l}{are for broad components with FWHM$\sim$1000~\kms. $*$APO data; $**$ MMT data.  Both are from \citet{IT09}. } \\
\end{tabular}
\end{table*}

\HeI\ emission line fluxes of $\lambda$4471 and $\lambda$5876 (the only lines
for which we have sufficiently good detections) appear the strongest at the LBV
discovery epoch (2008.01.11), when the LBV was $\sim$2 magnitudes fainter with
respect of its maximum on 2008.03.28. Only line \HeI(4471) was within this
'maximal phase' spectrum range (see \citet{IT09}).
Its flux was an order of magnitude lower than in the LBV discovery spectrum.
For somewhat fainter LBV level in two other dates, \HeI(4471) flux was also
low, but probably a factor of $\sim$1.5 higher than in the maximum.
For \HeI(5876) line, there are only two spectra with the apparent LBV
contribution. So, the only firm statement we can make is that the flux of
\HeI(5876) line is down by a factor of $\sim$8$\pm$2 between epochs
2008.01.11 and 2009.01.21, while the LBV $V$-band luminosity is up by a
factor of four.
This, at first look, is consistent with the run of \HeI(4471) line flux.

The presence of emission He{\sc ii}(4686) line is an indicator of the hot LBV state
characteristic of minimums in LBV lightcurves, when a star approaches WR-type
parameters (e.g. \citet{Sholukhova11}). In this state, the EW(He{\sc ii}~4686) can
reach $\sim$10~\AA.
We did not detect He{\sc ii}(4686) line in any of our spectra, and there
were no traces of this line in \citet{IT09} spectra. The common upper
limit (2$\sigma$) on the line flux corrected for the loss on slit on our
spectra is $\sim$0.2 $\times$ 10$^{-16}$ erg cm$^{-2}$ s$^{-1}$.
Since the contribution of the LBV continuum in our spectra during years
2004--2005 to the light of the underlying \HII\ region was less than
several percent, this line should be below the detection level.

The variations of the LBV Hydrogen Balmer line intensities are not easy to
follow since they require deblending of the composite line profiles (\HII\
region plus LBV) and extraction of their broad components. They also need to correct
for the loss of the LBV light on the slit. Both their width and the
wind terminal velocities estimated from P-Cyg profile remain the same within
the measurement uncertainties.

Unfortunately, for H$\alpha$ we have only two spectra with the interval of
three weeks (2008.01.11 at BTA and 2008.02.02 from \citet{IT09}). But
during this time the LBV $V$-band luminosity had increased by a factor of four.
The flux of H$\alpha$ dropped by factor of 2.2$\pm$0.5. The similar drop
of H$\beta$ flux was registered. However, the subsequent variations of H$\beta$
flux appeared to be less predictable. In particular, in the maximum LBV
phase (2008.03.28), F(H$\beta$) was close to that of the LBV discovery date,
when the LBV was $\sim2^m$ fainter. In 10 months after the maximum, when the
LBV $V$ band luminosity dropped by $\sim$40~\%, F(H$\beta$) increased by a
factor of 1.5.
Surprisingly, in comparison to the similar LBV level on 2008.02.02-04,
F(H$\beta$) became 2.5 times higher.
The estimates of fluxes of H$\gamma$ line are of smaller S-to-N, but in
general agree with variations of H$\beta$ flux.

\section{H$\alpha$ shells and the super giant shell}
\label{sec:shells}

\subsection{The Super Giant Shell}
\label{sec:SGS}

The most prominent H$\alpha$ emission in DDO~68 is related to the outer region
at the northern edge, named in PKP05 the `Northern ring'. It consists
of at least five SF regions with the range of H$\alpha$ luminosities and
ages of 3--7 Myr as inferred from the comparison of their $EW$(H$\beta$)
with predictions of temporal behaviour of this parameter in the Starburst99
models  \citep{Starburst99}.
The total H$\alpha$ flux from DDO~68, as found in PKP05, is
1.73$\times$10$^{-13}$ erg~s$^{-1}$~cm$^{-2}$ which corresponds to the
luminosity
$L$(H$\alpha$) = 3.38$\times$10$^{39}$ erg~s$^{-1}$. The major part of this
luminosity
($\gtrsim$ 80\%) originates in the 'Northern Ring' (see Fig.~1, right panel.
Continuum is not subtracted.)
In PKP05, these SF regions appeared barely resolved and were called
'Knots' with the respective numbers. In the HST image they all are
extended and well resolved. Therefore it will be more suitable to call
them Star-Forming (SF) regions, keeping the same numbers.

In fact, these five SF regions, well visible in the H$\alpha$ image, are
complemented by the 6-th object, the compact young stellar cluster (YSC) (age
of $\sim$22~Myr, PKP05), which is seen in the HST H$\alpha$ and broad band
images as an almost star-like source (FWHM($V$)=0\farcs14, compared to
FWHM=0\farcs10 for the ACS Point Spread Function (PSF)).
Thus, the intrinsic FWHM(YSC) $\sim$0\farcs10, or
$\sim$6~pc. The compact core of the YSC is surrounded by a diffuse halo of lower
contrast and with the full extent of $\sim$0\farcs3.
The overall size of the 'Northern ring' is
$\sim$15\arcsec$\times$18\arcsec\ that corresponds to the linear 'diameter'
of $\sim$1.1 kpc.

Similar large structures, Super Giant Shells (SGS) and rings have been known
for a long
time both in our Galaxy and in other nearby galaxies. They are well identified
in particular in the suitable angular and velocity resolution \HI\ maps of
nearby galaxies through the prominent `holes' in the gas distribution and
its expansion in the compressed walls around the holes. They are produced, most
probably, by the cumulative action of multiple supernovae (SNe) and of massive
star winds in the intense SF episodes (see e.g. review by
\citet{Elmegreen98}).

The ambient gas compressed by such shells and their shock fronts can, at
certain conditions, appear unstable and collapse, leading to the sequential
induced SF episodes. In particular, SGSs with the maximum sizes
of $\sim$1--1.5~kpc are found in several galaxies. Well-studied objects
are IC~2754 (e.g., \citet{Egorov14} and references therein) and Holmberg~II
\citep{Puche92}, both from the nearby group M81. Expansion velocities of
such SGSs vary across a rather wide range. For example, in Holmberg~II,
according to
\citet{Puche92}, $V_{\rm shell}$ are of the order of 10~\kms\ and their
estimated characteristic expansion times (0.6~$R/V_{\rm shell}$) are of
35--60~Myr. For the SGS in
IC~2754, with the overall size of 0.8 kpc, \citet{Egorov14} find
$V_{\rm shell} =$ 25~\kms\ and the expansion time of $\sim$10~Myr.

\begin{figure*}
  \centering
\includegraphics[angle=0,width=7.5cm, clip=]{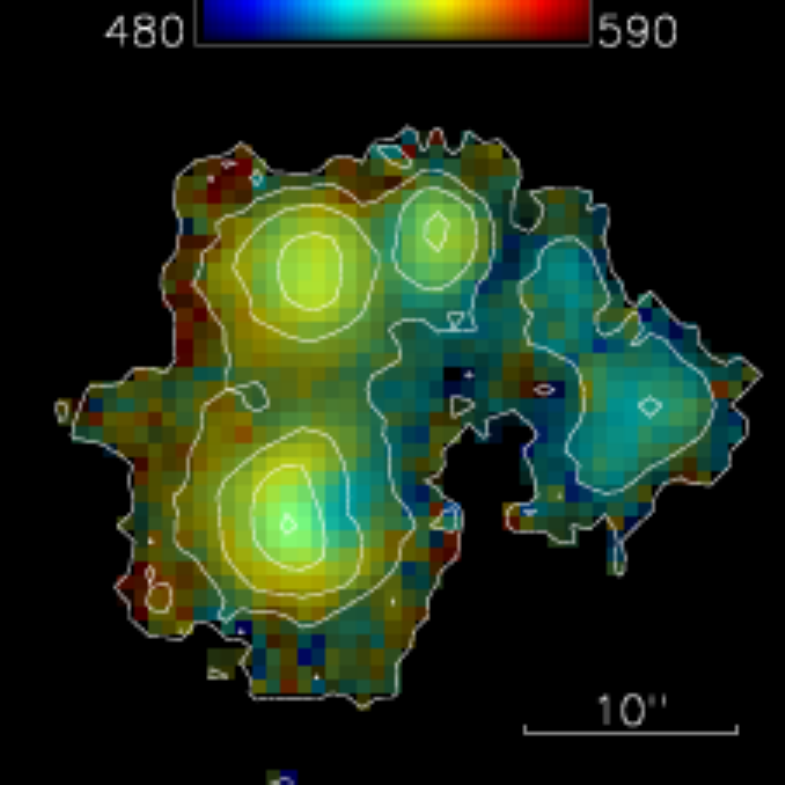}
\includegraphics[angle=0,width=7.5cm, clip=]{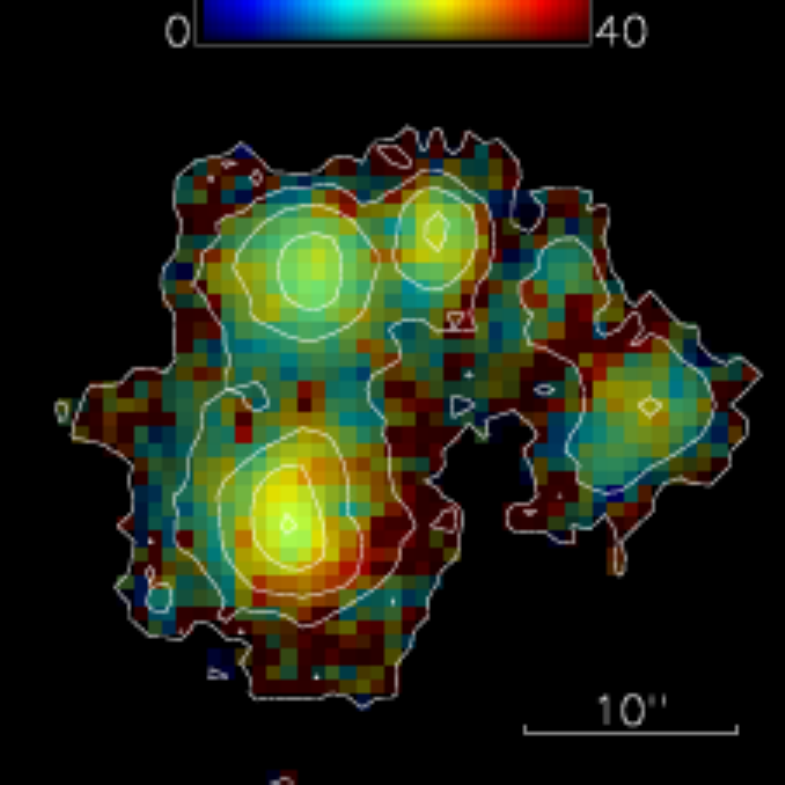}
\caption{The FPI observations of the supergiant shell: the maps of
H$\alpha$ velocities (left panel) and velocity dispersion (right panel).
The pixel brightness is scaled according to the H$\alpha$ intensity, the
colors correspond to velocities (the scales on the top are in \kms). The
overlaid contours show the logarithmic scaled H$\alpha$ intensity.
}
\label{fig:FPI}
 \end{figure*}

\subsection{SGS expansion velocity}
\label{ssec:SGSexpanvel}

In Fig.~\ref{fig:FPI}, we show the ionized gas velocity and velocity
dispersion maps derived from fitting of the H$\alpha$ profiles in the FPI
data cube.
The velocity field reveals a significant gradient of the
line-of-sight velocities at the East--West direction, i.e.
roughly perpendicular to the galaxy disc major axis.  We
accept the disc orientation parameters (inclination $i=65\degr$ and the
line-of-nodes position  angle $PA_0=17\degr$) adopted from the comparison
of \HI, \HII, and photometric data \citep{Moiseev2014}. According to this
paper, both H$\alpha$ and \HI\ rotation curves have a flat shape at
the radial distances corresponding to the location of the SGS. Therefore
variations of the line-of-sight velocity related to the disc rotation
should be negligible.
A possible explanation
of the observed velocity gradient is a radial expansion of the SGS.

\begin{figure*}
  \centering
 \includegraphics[angle=0,width=7.5cm, clip=]{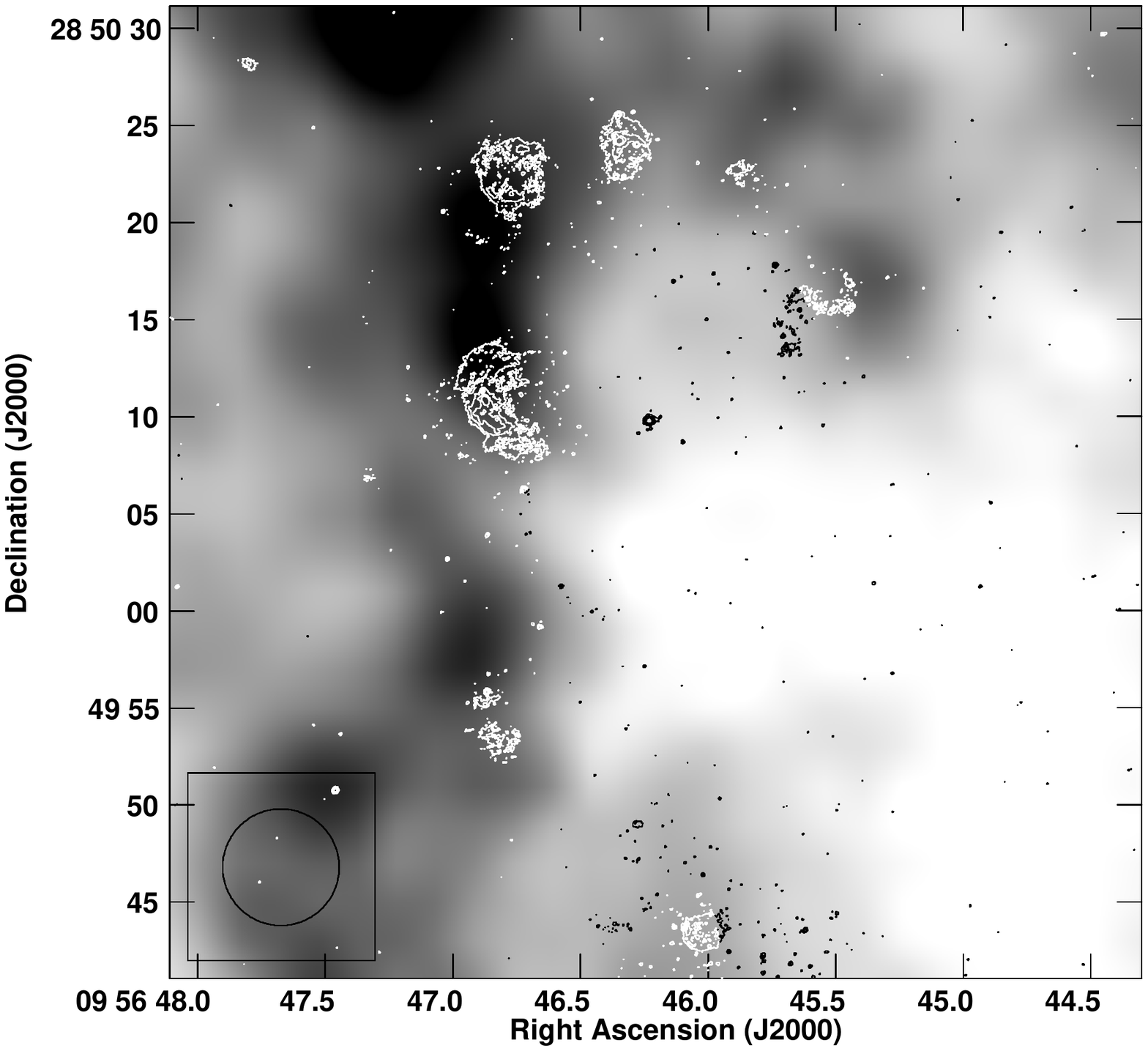}
 \includegraphics[angle=0,width=9.0cm, clip=]{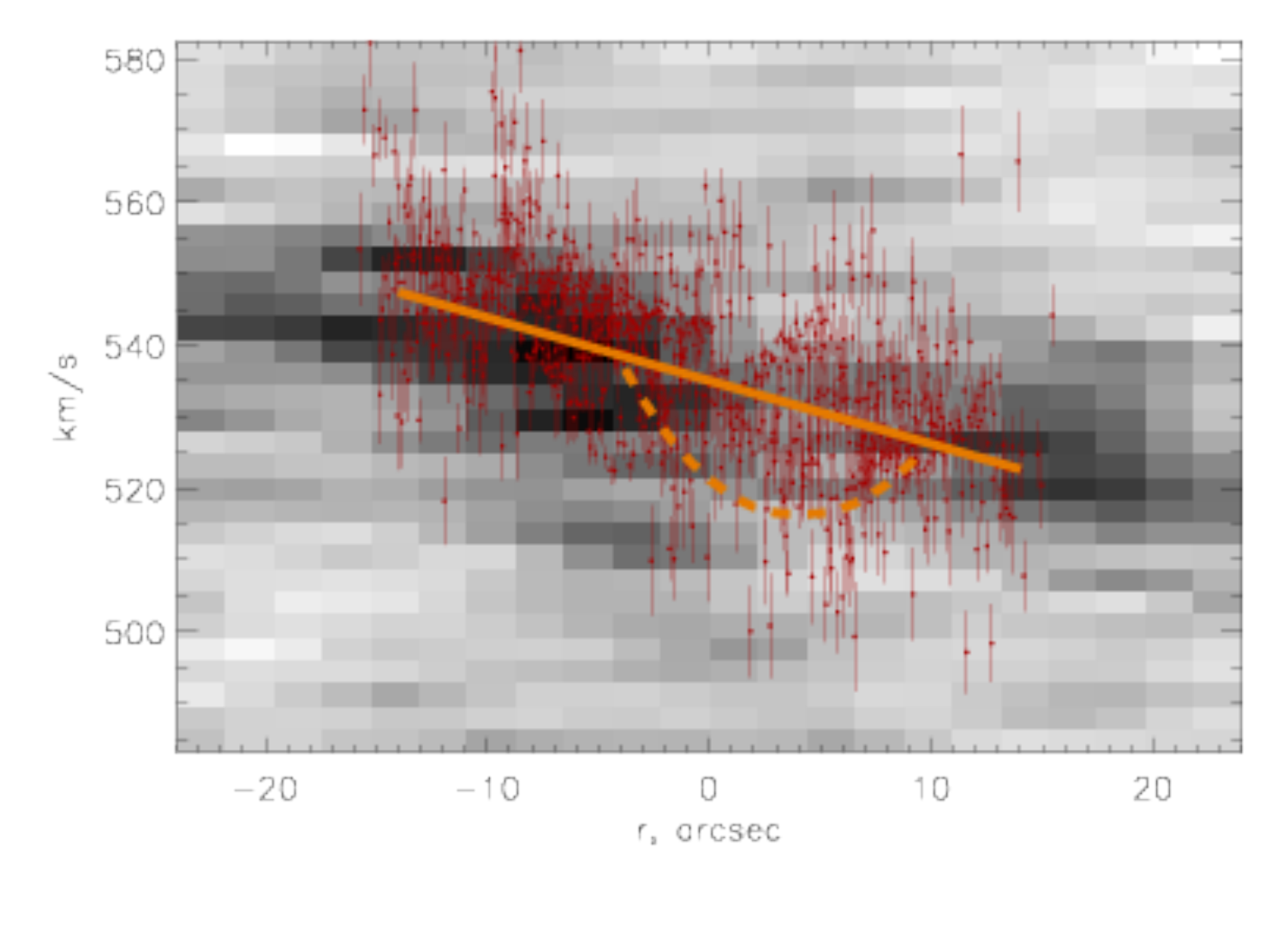}
  \caption{
{\bf Left panel:} the HST H$\alpha$ image in white contours overlayed on the
GMRT \HI-map MOM0
(flux density, in grey scale) of DDO~68 with the beamsize of
$\sim$5\farcs5 (shown as a circle in the bottom-left corner).
The SGS is outlined by bright H{\sc ii} regions (as in Fig.2 left)
north of the region devoid of \HI\ (central part in white).
There is  '\HI\ hole' -- region of the strong decrease of \HI\ density,
(3--5 times relative to that on the ring borders) in the central part of
the SGS, while the density is high enough along the perimeter of the SGS, and
in particular, close to the regions of current SF well seen in H$\alpha$.
N is up, E is to the left.
{\bf Right panel:}  P--V diagrams of \HI\ (grey scale) and H$\alpha$ (red
points with error bars) velocities across the SGS region with the width of
the 'lane' of 18\arcsec. Solid straight line shows the linear fit for P--V
diagram. The dashed curve models the line-of-sight velocities of
the approaching part of the shell, from which we estimate
$V_{\rm SGS} \lesssim$ 15~\kms.
See details of description in text of Subsec.~\ref{ssec:SGSexpanvel}.
}
	\label{fig:HImap}
 \end{figure*}

We estimated  $V_{\rm shell}$ using the neutral and ionized gas
velocity distribution along the broad cross-section including all \HII\
regions in the SGS with $PA=107\degr$ that is parallel to the disc minor axis.
The respective  P--V diagrams of both \HI\ (grey scale) and H$\alpha$
velocity fields  across the region of the SGS and its surroundings are
shown in the right panel of Fig.~\ref{fig:HImap}. As one can see, despite
the large scatter of H$\alpha$ data, they match each other rather well.
In regions near the major axis of a regular rotating flat disc, we expect
a zero gradient of the line-of-sight velocities across the direction parallel
to the minor axis of a galaxy.
Therefore, the apparent gradient in this P--V diagram with the total
velocity range of $\sim$30~\kms\ looks unexpected and implies
disturbance of the regular disc rotation.

The centre of this P--V feature coincides with the position of the SGS
and \HI\ hole, giving us an additional argument in favour of  the
hypothesis that the observed velocity gradient is caused by an expanding
shell. The resulting values for the line-of-sight projection of the expansion
velocity at the SGS borders ($r= $ 14~arcsec) fall in the range of
$V_{\rm obs}=10-15$~\kms\ that corresponds to
$V_{\rm shell}=V_{\rm obs}/\sin i=13.5$~\kms\ with the uncertainty
of 2.5~\kms.

An alternative explanation of the velocity gradient in Figs.~\ref{fig:FPI}
and ~\ref{fig:HImap} lies in the tidally-induced streaming motions caused by
the recent merger. However, we can present  at least two arguments
against its  tidal origin:

\begin{enumerate}
\item Indeed, the kinematical
   major axis derived from the galaxy velocity field
 is twisted with position angle deviated from $PA_0$. In the region of the
 SGS  the kinematic position angle $PA_{\rm kin} \approx 30\degr$
\citep[see the velocity field in Fig.~2 by][]{Cannon14}.
However, when we construct the P--V diagram along the direction orthogonal
to this `local' major axis direction  (i.e. along $PA \approx 120\degr$), the
linear velocity gradient across the SGS is still detected, while it falls
to near zero value outside the SGS region ($r>30-40$ arcsec).
This fact indicates  the local origin of the \HI\ velocity deviations in this
region, seemingly related to the radial expansion in the galaxy disc.
 
\item  According to Fig.~\ref{fig:HImap}, the H$\alpha$ velocities coincide
very well with the linear fit to the P--V diagram on the borders of the SGS,
while a blue excess is observed in the SGS centre up to 10--20~\kms\.
The most reasonable interpretation of this feature is associated with the
approaching part of the ionized gas expanding shell, a half of the classical
`velocity ellipse' \citep[e.g. Figs.~6 and 10 in][]{Egorov14}.
The dashed curve in Fig.~\ref{fig:HImap}  fits this feature
with $V_{\rm shell} \le$15~\kms. Moreover, some pixels in \HI\  \mbox{P--V}
diagram in this region appear to trace both the approaching and receding
sides of the shell,
but, unfortunately,  the signal-to-noise is too low to make unambiguous
conclusion. The most of \HI\ gas in this region is expected to be ionized
and/or swept out.
\end{enumerate}

Summarizing, we see that the two independent estimations: the first one based
on the velocity gradient itself (the radial expansions in the disc) and the
second one derived from the H{\sc ii} velocities (expansion in the 'vertical'
direction) give the close values of the $V_{\rm shell}$.

\subsection{SGS expansion time}

Adopting the current radius $R_{\rm shell}$ as 0.55~kpc and
$V_{\rm shell} =$ 13.5$\pm$1.5~\kms, we derive the characteristic expansion
time of 0.6~$R_{\rm shell}$/$V_{\rm shell} = 24\pm5$~Myr.
This value indicates how long ago did the multiple SN explosions begin
in the process of the intense SF episode in the central part of SGS, which provided
the thermal and mechanical energy for the \HI\ shell expansion.
If we add 3.5 Myr needed for an instantaneous star formation episode to
explode its first SNe, we derive that the time since the SF began in the
center of the SGS is  27.5$\pm5$~Myr.

\begin{figure*}
  \centering
\includegraphics[angle=-0,width=7.5cm,clip]{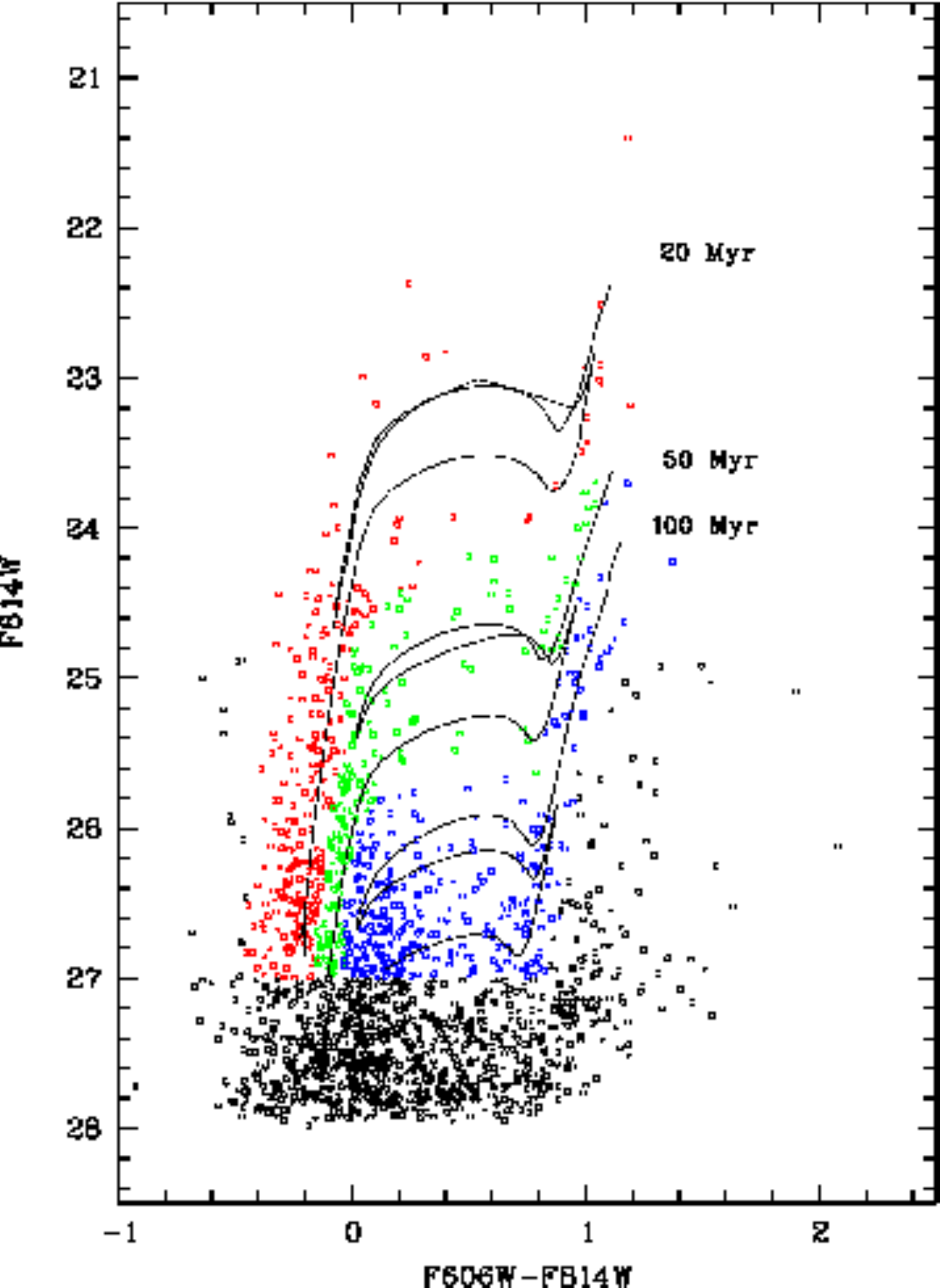}
\hspace{0.5cm}
\includegraphics[angle=-0,width=7.5cm,clip]{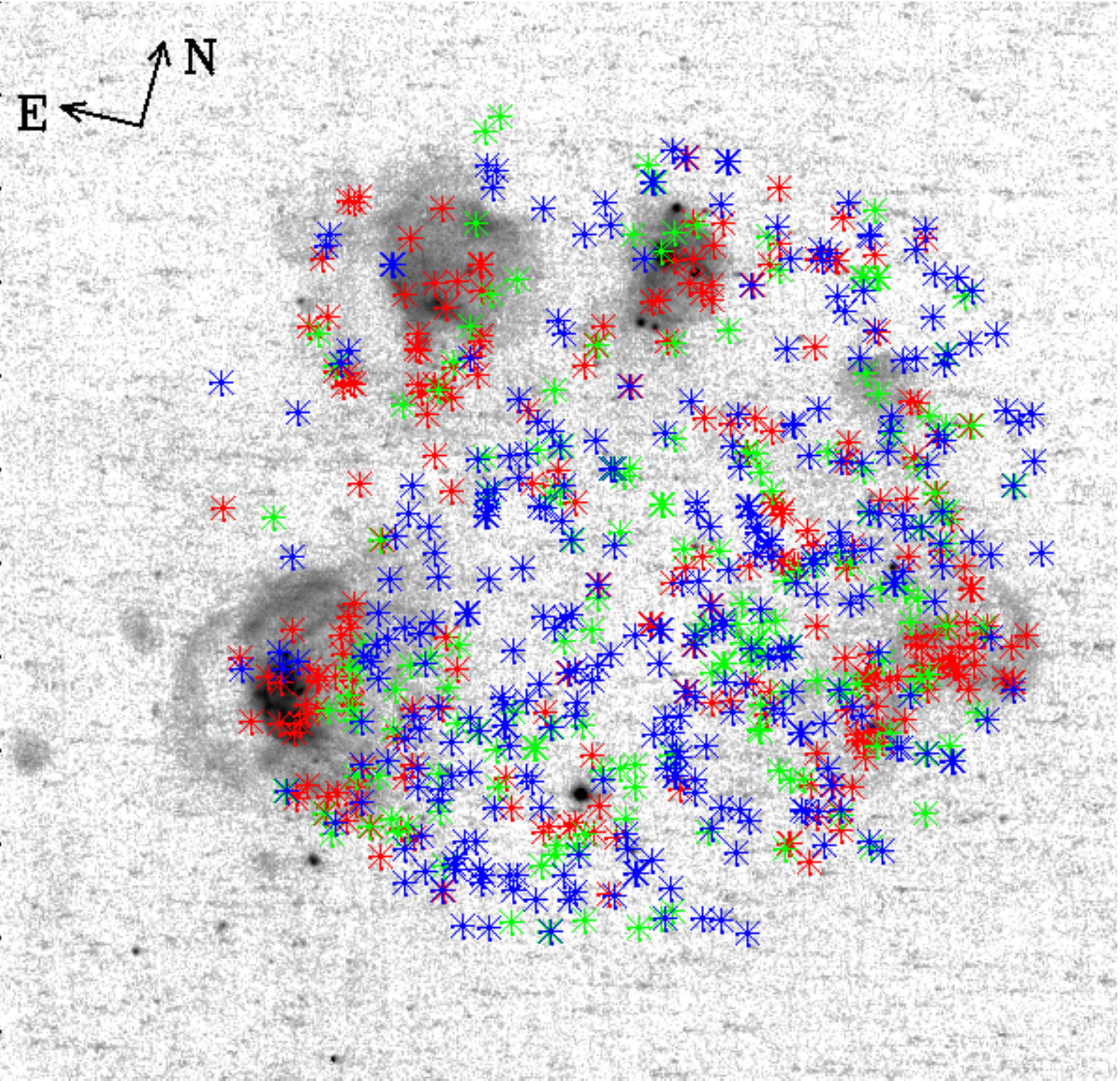}
  \caption{
{\it Left panel. } Color-magnitude diagram (CMD) for the stars
within the radius of 11.5~arcsec (0.7~kpc) from the SGS centre.
The stellar populations
of different ages are marked with different colours.
Stars younger than $\sim$20 Myr are shown in red,
those with ages between $\sim$30 and $\sim$60 Myr are
marked in green and stars with ages between $\sim$60 and $\sim$100~Myr
are in blue. We do not consider the stars fainter than $\sim$27$^m$
to avoid the higher photometric uncertainties areas.
{\it Right panel. }
Spatial distribution of the stellar population in the
region of the SGS overlaid on the {\it HST} H$\alpha$ image.
The colours of individual stars (asterisks) correspond
to those in the CMD in the left panel.
The plot gives a clear evidence that the stars younger than
$\sim$30 Myr reside predominantly near the boundary of the
SGS, concentrating in the related young associations.
The stars with ages of 30--50 Myr and 50--100 Myr are
distributed rather evenly across the selected area.
}
	\label{fig:CMD_iso}
 \end{figure*}

An independent indication for the delay between the SF episode near the
central part of the SGS and on its perimeter regions is shown in
Fig.~\ref{fig:CMD_iso}. In the left panel, the CMD with the overlayed
isochrones shows with red, green and blue colour, the stars respectively
with the ages less than $\sim$20~Myr,
and in the ranges 30--50~Myr and 50--100~Myr.
In the right panel we show the spatial distribution of these age groups with
respect to the SGS geometry as traced by the star-forming regions in the
{\it HST} H$\alpha$ image (in grey scale).
The `young' population (red asterisks) is highly concentrated
along the SGS perimeter, grouping to the SF `Knots' mentioned earlier.
The stars with larger ages are distributed more evenly, which probably
indicates a more homogeneous star formation on the time scale of
50--100~Myr and overlapping of different layers due to projection
effects.
Also, this does not exclude their origin near the
SGS centre, since due to velocity dispersion of $\sigma_V \sim$ 10~\kms\
they are expected to travel during the lifetime to distances of
several hundred parsec.

In the context of similarity of this SGS to other known objects,
the SGS in DDO~68 has the size and
expansion velocity well comparable with those in galaxy Holmberg~II.
As \citet{Egorov14} show in their Fig.~1 for the SGS in IC~2754 in the
H$\alpha$ HST image, there are several clearly visible SF regions along the
SGS borders. Moreover, they are related to the secondary H$\alpha$
arcs/shells tracing the recent induced SF episodes on the SGS wall. These
shells have sizes of $\sim$50 to $\sim$300 pc and characteristic expansion
times of $\sim$1--4~Myr.

\subsection{SGS morphology}

Distribution of star-forming regions along the walls of a SGS described
in the previous section
hints at induced star formation related to the gas instability in the compressed shell and to
collapse of matter swept by the expanding shell of the previous powerful
starburst and related multiple supernova (SN) explosions
\citep[e.g., review by][]{Elmegreen98}.

The `Northern ring'  delineates a supershell with the large \HI\ hole.
This is clearly seen in the \HI-map of Fig.3 (right) of \citet{Ekta08}. In
Fig.~\ref{fig:HImap} (left) we present a fragment of the GMRT \HI-map with the
beamwidth FWHM=5\farcs5 superimposed on the {\it HST} H$\alpha$ image of the
SGS and its surroundings. One can see the strong decrease of \HI\ density in the
interior of the SGS with respect to that at its periphery where the regions
of current and recent SF shine.

Unfortunately, the available \HI\ data have rather low spatial
resolution. Thus, using only \HI\ density and velocity distributions, we
can not conclude unambiguously whether the observed neutral gas
features are `local' (that is directly related to the stellar feedback in
the SGS region), or the \HI\ hole is a part of larger-scale \HI-structures
related to a strong tidal disturbance.
However, the whole set of the observed
features (the \HI\, hole, the radial expansion of neutral and ionized gas,
a relation  between our estimates of the expansion time and stars ages, etc.)
is very similar to those for SGSs observed with a higher spatial resolution
in some of the aforementioned nearby galaxies.
This allows us to believe  that our main suppositions are true.

\subsection{H$\alpha$ shells and their properties}

In the BTA H$\alpha$ image of this region in PKP05 (their Fig.~1, seeing
$\beta \sim$ 1\farcs7) most of \HII\ regions are only barely resolved and their
structure is difficult to assess.
The {\it HST} H$\alpha$ image, with $\sim$20 times higher angular
resolution and the higher dynamic range, allows one to distinguish multiple
ionized gas arcs/rings  with the typical sizes of 1\arcsec\ to
$\sim$4\arcsec, corresponding to linear diameters of $\sim$60 to
$\sim$250~pc related to the regions of current or recent star formation.
The larger the size, the lower the surface brightness (SB) of an arc/ring.
The similar H$\alpha$ morphology was observed in many nearby galaxies,
including, e.g., the Local group late spiral M33 \citep{Courtes87} with the
range of ring diameters of 40 to 280 pc.

One can see the large diversity in parameters of the second generation SF
episodes and their descendants -- from a rather loose complex related to
Knot~4 through more powerful ones with the dominant luminous compact
`knots/stars' (Knots~1 and 3) to the very compact star complex identified as
the `Young Stellar Cluster' (Knot~5). One of the reasons for this broad
diversity is probably related to the ambient gas inhomogeneity and lumpiness.
The  most compact newly formed stellar aggregate, YSC, had formed on the
southern part of SGS front,
bordering the region of very low gas density as seen in Fig.~\ref{fig:HImap}.

The brightest SF region Knot~1 appears as a collection of fragments of seven
H$\alpha$ arcs with the SB (surface brightness) decreasing with the arc/ring
radius. The two
brightest and smallest arcs/rings with the characteristic radii of 0\farcs8
(48~pc) and 1\farcs0 (60~pc) look like they are crossing (in projection?) each other
(with the separation between their centers of $\sim$1\arcsec) and appear
close to the group of at least 15 bright  compact SF subregions either
unresolved ($<$0\farcs1, or $\lesssim$6~pc) or with the typical sizes of
6--12~pc. At least a half of them 'stick' to the arcs or appear rather close
to them. Five other lower SB arcs with radii of $\sim$2\arcsec\
($\sim$120~pc) to $\sim$4\arcsec\
($\sim$250~pc) encompass this region. Their centres of expansion are
displaced each from other up to
several arc seconds (or more than 100--150 pc in projection).

Unfortunately, our FPI data (the final seeing 1\farcs8) barely resolve
the spatial structure of this region. Nevertheless,  the H$\alpha$ kinematics
maps reveal some interesting details in Knot~1 (Fig.~\ref{fig:FPI}). Namely,
there is a difference of the line-of-sight velocities up to 20--22~\kms\
from the centre to the region border as well as an increase of the ionized
gas velocity dispersion on the border of Knot~1 where the low SB arcs appear
on the HST image (see above).
The mean velocity dispersion in these regions is about 29~\kms\, which
corresponds to  $\sigma_{\rm cor}=27.0\pm0.5$~\kms\ after correction
for the natural width of the emission line and its thermal broadening
\citep[see eq. (1) in][]{MoisLoz2012}.
If we accept $V_{\rm shell}\approx\sigma_{\rm cor}=$ 27~\kms\ and
$R =$ 5\arcsec\ (310 pc), then the dynamical age of Knot~1 emission shell
is about 7 Myr.

As the estimates of the possible shell ages, presented above,
imply, this picture
indicates the continuing active SF on the time scale of a few Myr at
least in seven different sites within the region of the characteristic size of
100~pc. While this sequence of SF
events was apparently triggered by the gas instabilities in the compressed
supershell, we also see the secondary triggered SF knots situated along the
brightest H$\alpha$ arcs.

SF region Knot~2 is much fainter in H$\alpha$. There are two brighter compact
SF subregions and several fainter ones in this region. Most of them are
either close
to or 'stuck' to the arcs of 2 shells with radii of $\sim$2\arcsec. There
is a hint of a very low SB arc separated at $\sim$4\arcsec\ to the E.

SF region Knot~3 displays one almost round shell with the size of
$\sim$1\farcs9 by 1\farcs7. Several compact subregions with sizes of
$\lesssim$0\farcs2 are situated close to the shell or to the adjacent arc
to the N. The brightest
one sits right on the shell. This subregion contains the LBV (DDO68-V1).
In fact, as its light profile shows, it is well fit by the PSF. So, the
star-like LBV with its expected parsec-scale shell ($\sim$0\farcs015 at
the distance of DDO~68) dominates the light of this subregion.

There are two faint arcs related to SF region Knot~4 which encompass four
compact SF subregions. Their diameters are $\sim$3\arcsec\ and $\sim$1\farcs7.
One more faint H$\alpha$ nebulosity with the size of $\sim$1\farcs7 is
apparent along the same Northern supershell between SF Knots~3 and 4.
But no compact subregions are detected within this nebulocity.
The FPI ionized gas velocity dispersion map demonstrates a peak of velocity 
dispersion in the center of this SF region, which is to be expected in
the case of a simple  expanding shell \citep{MunozTunon1996, MoisLoz2012}.
Using the same techniques as above for Knot~1, we estimated the corrected
velocity dispersion $\sigma_{\rm cor}=23.4\pm0.9$~\kms\ and the respective
dynamical age of Knot~4 of 4~Myr
for the region radius $R =$ 2\farcs5 (155 pc).

SF region Knot~5 is a 'closing unit' of the chain, forming the perimeter of
the SGS. It is related to the young stellar cluster with the estimated age
of 22 Myr (see PKP05).
It has an almost unresolved core ($\sim$0\farcs2) and the surrounding
halo with the overall size of $\sim$0\farcs6. This is, probably, evidence to
the remnant of the related  \HII\ region of low excitation.
It is curious that region Knot~5 is the only one where no traces of
H$\alpha$ arcs are visible. This either hints at a much lower density of the
ambient gas with respect to the other regions,
or at their premature formation and dissolving, or both.

All this complex morphology seen as a combination of arcs from the expanding
shells and compact bright objects with the linear sizes of 15 to 50 pc,
situated preferably very close to the arcs  brightest edges, apparently
indicates to the induced hierarchical star formation over the course of the last
$\gtrsim$ 30~Myr.

\setcounter{qub}{0}

\begin{table}
\caption{DDO 68 brightest blue and red supergiants in and
near the SF regions with the record low values of O/H.}
\label{t:lum.list}
\begin{tabular}{rrrrrrl}
\hline
N    & $X$    &  $Y$ &  $V$ & $(V-I)_{\mathrm 0}$& $M_{\rm V}$& \HII-reg. \\
     & px     &  px  &  mag & mag  &  mag       & Numb. \\
\hline
\qq  & 3455.8 & 622.9& 20.05&--0.10&--10.52&  3, LBV \\
\qq  & 2933.4 & 925.2& 21.47&  0.33& --9.11&  6 \\
\qq  & 3447.8 & 605.9& 22.72&  0.30& --7.86&  3 \\
\qq  & 3076.0 & 923.8& 22.93&  0.15& --7.65&  6: \\
\qq  & 3295.5 & 504.4& 23.00&  1.56& --7.58&  4 \\
\qq  & 3222.6 & 518.3& 23.10&  0.06& --7.49&  4 \\
\qq  & 3139.7 & 651.2& 23.30&  0.40& --7.28&  5 \\
\qq  & 3243.8 & 476.5& 23.35&  0.13& --7.24&  4 \\ 
\qq  & 3461.8 & 615.7& 23.38&  0.50& --7.20&  3 \\
\qq  & 2986.9 & 852.1& 23.41&--0.11& --7.17&  6 \\ 
\qq  & 2927.5 & 938.2& 23.42& -0.16& --7.16&  6 \\
\qq  & 3245.1 & 493.5& 23.47&--0.10& --7.11&  4 \\
\qq  & 3213.6 & 825.5& 23.81& -0.09& --6.77&  1 \\
\qq  & 3243.9 & 475.6& 23.95&--0.13& --6.63&  4 \\ 
\qq  & 3302.7 & 611.3& 23.96&  1.41& --6.62&  4: \\
\qq  & 3301.0 & 568.9& 23.98&--0.07& --6.60&  4: \\
\qq  & 2985.9 & 850.1& 24.09&  0.21& --6.49&  6 \\ 
\qq  & 3236.4 & 836.2& 24.13&--0.36& --6.45&  1 \\
\qq  & 3393.2 & 749.1& 24.13&--0.21& --6.45&  2 \\
\qq  & 3255.3 & 493.9& 24.15&--0.18& --6.43&  4 \\
\qq  & 3158.5 & 693.0& 24.22&  0.25& --6.36&  5 \\
\qq  & 3372.9 & 468.4& 24.26&  0.24& --6.32&  4 \\
\qq  & 3256.2 & 497.2& 24.27& -0.22& -6.31 &  4  \\
\qq  & 3480.9 & 599.8& 24.30&  1.32& -6.28 &  3  \\
\qq  & 3223.5 & 798.4& 24.32& -0.09& -6.26 &  1  \\
\qq  & 3205.6 & 830.8& 24.35& -0.70& -6.24 &  1  \\
\qq  & 3286.0 & 468.3& 24.35& -0.16& -6.23 &  4  \\
\qq  & 3179.5 & 622.7& 24.30&  0.22& -6.23 &  5  \\
\qq  & 3444.8 & 728.7& 24.36&  1.41& -6.22 &  2  \\
\qq  & 3179.6 & 846.4& 24.36& -0.15& -6.22 &  1  \\
\qq  & 3481.9 & 614.8& 24.39& -0.54& -6.19 &  3  \\
\qq  & 3251.1 & 521.6& 24.40& -0.11& -6.18 &  4  \\
\qq  & 3210.2 & 518.4& 24.40& -0.52& -6.18 &  4  \\
\qq  & 3166.2 & 811.7& 24.41&  0.24& -6.17 &  1  \\
\qq  & 3439.9 & 605.8& 24.43& -0.18& -6.15 &  3  \\
\qq  & 3213.8 & 837.4& 24.46& -0.37& -6.12 &  1  \\
\qq  & 2883.0 & 925.1& 24.47& -0.28& -6.11 &  6  \\
\qq  & 3348.9 & 687.8& 24.48&  0.03& -6.10 &  2:  \\
\qq  & 3404.2 & 744.0& 24.49& -0.20& -6.09 &  2  \\
\qq  & 3181.3 & 821.8& 24.49&  1.41& -6.09 &  1  \\
\qq  & 3174.5 & 828.4& 24.49& -0.07& -6.09 &  1  \\
\qq  & 3156.2 & 669.4& 24.50& -0.02& -6.08 &  5  \\
\qq  & 2917.5 & 909.3& 24.50&  0.08& -6.08 &  6  \\
\qq  & 3256.3 & 741.8& 24.52&  0.55& -6.06 &  1  \\
\qq  & 2951.0 & 918.2& 24.53&  1.44& -6.05 &  6  \\
\qq  & 2883.1 & 930.4& 24.53& -0.16& -6.05 &  6  \\
\qq  & 3218.5 & 822.2& 24.54& -0.22& -6.04 &  1  \\
\qq  & 3385.0 & 496.4& 24.55&  0.07& -6.03 &  4  \\
\qq  & 3449.6 & 742.4& 24.56& -0.30& -6.02 &  2  \\
\qq  & 2909.9 & 949.5& 24.56& -0.23& -6.02 &  6  \\
\hline
\end{tabular}
\end{table}

\section[]{The list of the most  luminous extremely metal-poor star in DDO68}
\label{sec:liststars}

We have selected a number of the brightest stars from the colour-magnitude
diagram. The absolute
magnitudes of this subsample stars are limited by $M_{\rm V} <$ --6.0 mag.
In addition, the list is limited only by the stars whose positions
correspond to the regions of current or recent star formation with the
measured \HII\ region record low O/H ($Z \sim$Z\sunn/35). These regions are
well seen in
the HST/ACS H$\alpha$ image as ring/arc plus compact source complexes.
This additional selection criterion is applied in order to pick out
only
the most metal-poor massive stars. This is necessary, since in contrast to
the most dwarf galaxies with small metallicity variations, in DDO~68,
as mentioned in the Introduction, stars show very large range of metallicities.
This is presumably due to the merger
of two dwarfs with very different $Z_{\rm gas,stars}$.

\begin{figure}
  \centering
 \includegraphics[angle=-90,width=7.5cm, clip=]{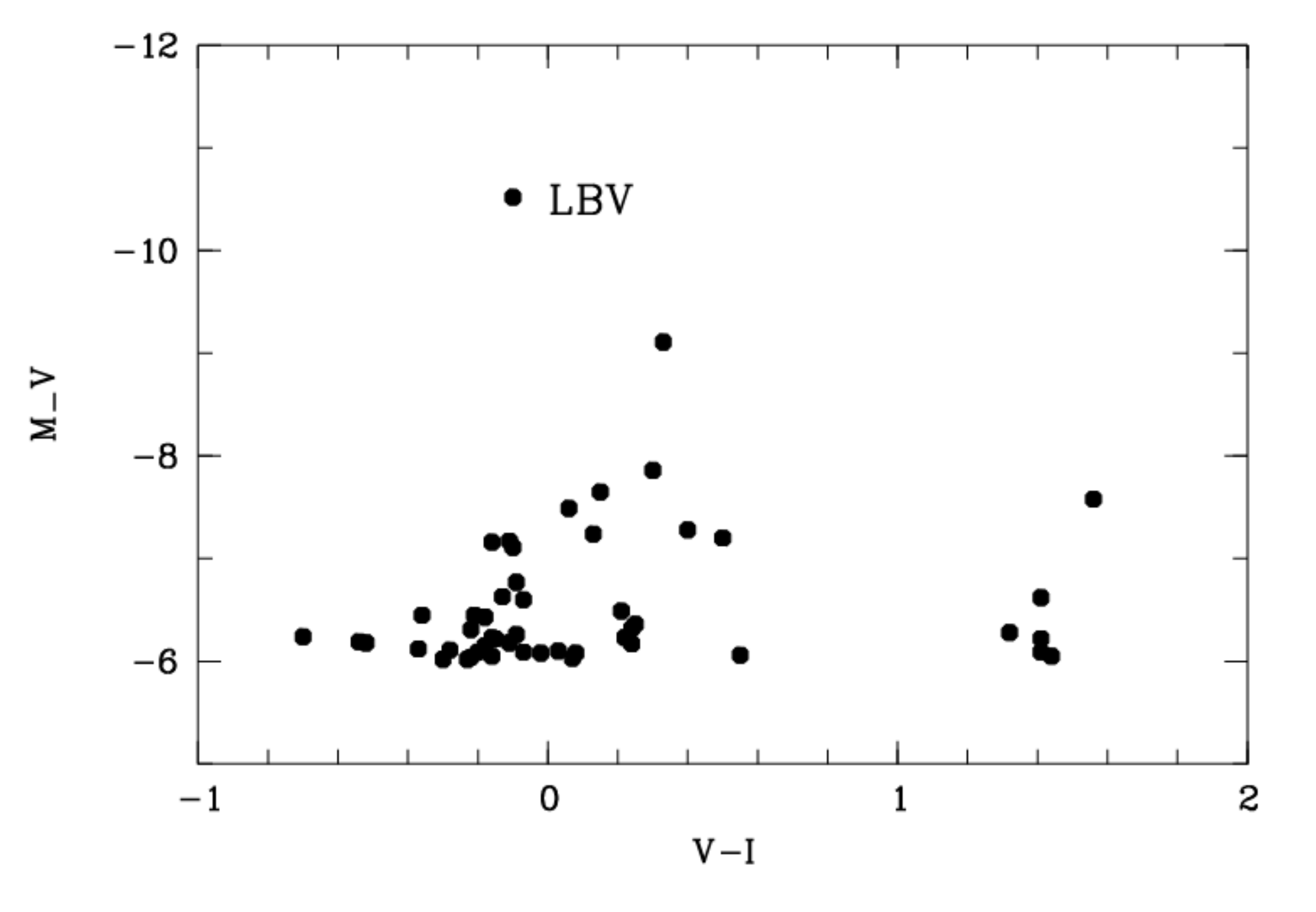}
  \caption{
The positions of 50 the most luminous extremely metal-poor stars of DDO~68
in the colour-magnitude diagram $M_{\rm V}$ vs $V-I$. They are mostly
diverse supergiants of spectral classes from O to M.
}
	\label{fig:CMD50lum}
 \end{figure}

The resulting list of such luminous stars is presented in
Table~\ref{t:lum.list}. It includes $X,Y$ coordinates of the stars in the
ACS/WFC2 frame, their uncorrected $V$ magnitudes and \mbox{$V-I$} colours
corrected for the Galaxy foreground extinction ($E(V-I)$ = 0.02$^m$ according
to \citet{SF2011}), their absolute magnitudes $M_{\rm V,0}$ and the number
of the SF region which this star is supposed to be associated to. We also
provide
the finding charts of all the selected stars arranged on their parent SF region
(see Fig.~\ref{fig:knot3}, \ref{fig:knot1_2_4_V}, \ref{fig:knot5_6_7_V}).
The amount of stars and the limiting luminosity
had been adopted in a rather arbitrary manner. The main criterion was to pick up a reasonable
number of the most luminous extremely metal-poor stars accessible for
spectral investigation with the next generation giant optical telescopes and
to make sure this list is not too large.

\begin{figure*}
  \centering
 \includegraphics[angle=0,width=7.5cm, clip=]{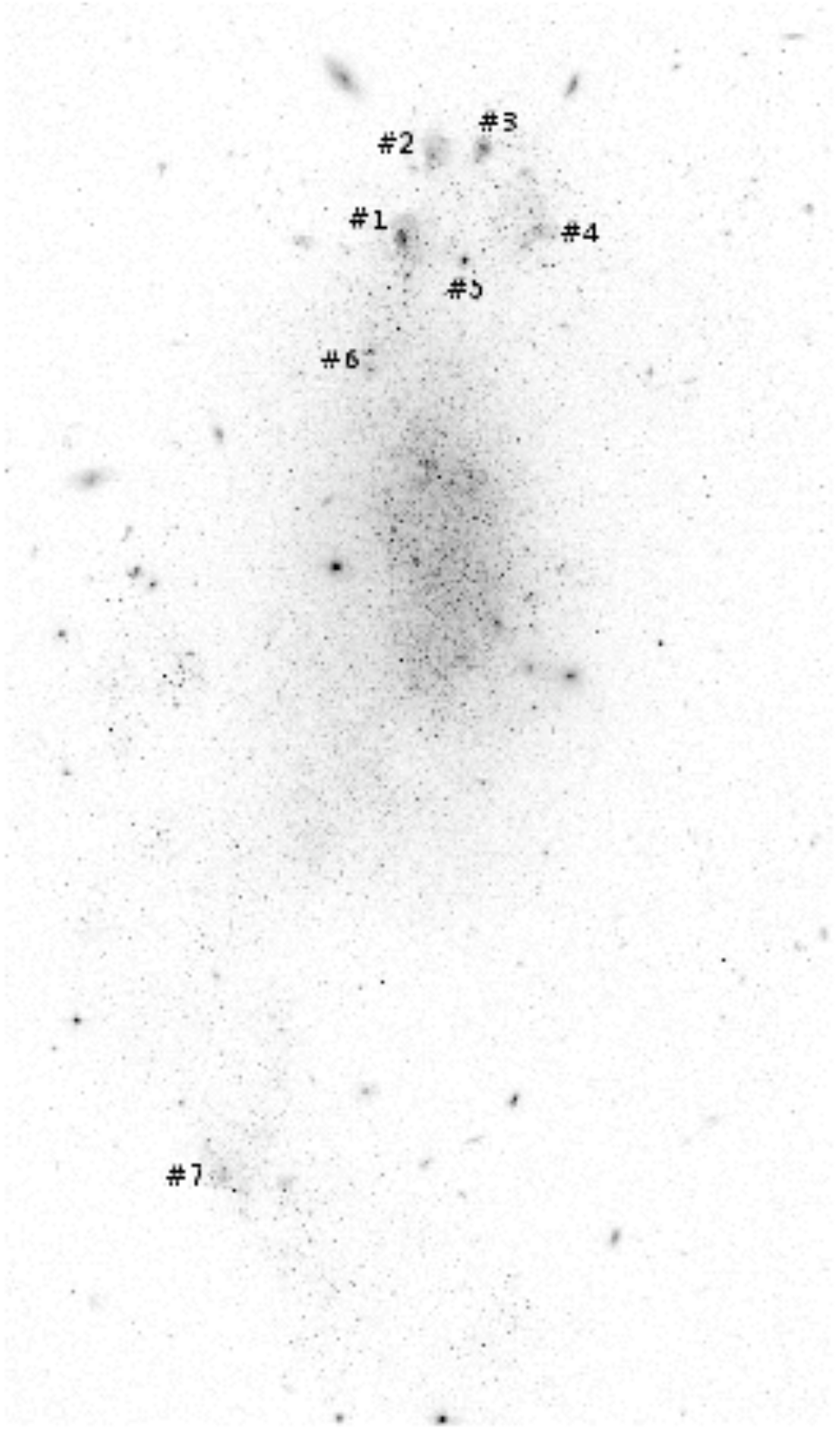}
 \includegraphics[angle=90,width=7.5cm, clip=]{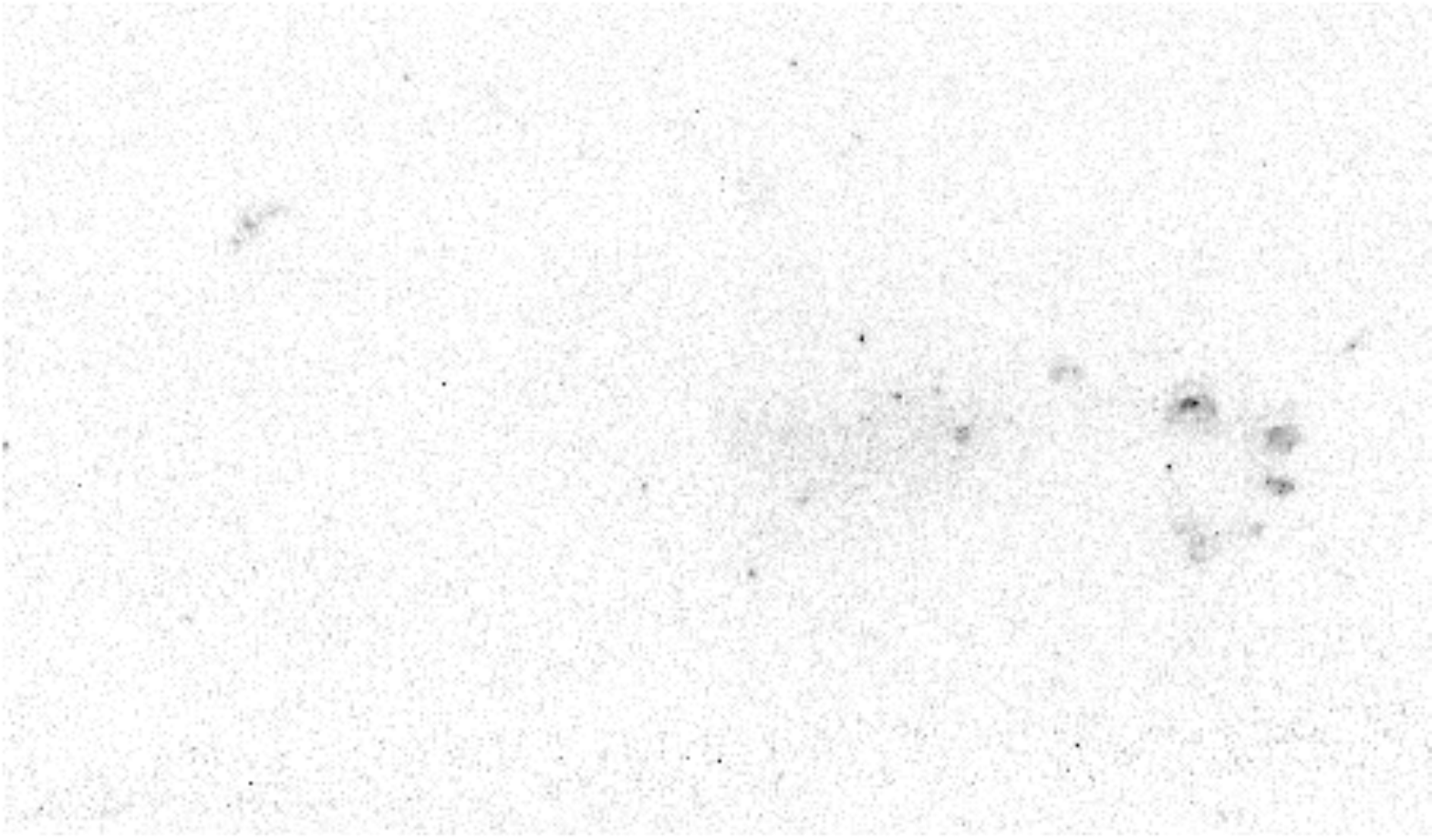}
  \caption{
{\bf Left panel}:
The {\it HST} image of the whole galaxy DDO~68 in $'V'$ band (F606W). Several
 regions of current/recent star formation are seen in the "Northern Ring"\
and one in the Southern tail (in this and the next images N is
approximately up ($PA=-14$\degr), E is to the left. See also arrows in other
figures).
{\bf Right panel}:
The {\it HST} image of DDO~68 in the narrow-band filter F658N (H$\alpha$)
showing  several regions of current/recent star formation, mainly in the
"Northern Ring"\ and the
southern tail, with a few knots in the main body.
}
	\label{fig:DDO68all}
 \end{figure*}

\begin{figure*}
  \centering
\includegraphics[angle=0,width=7.5cm, clip=]{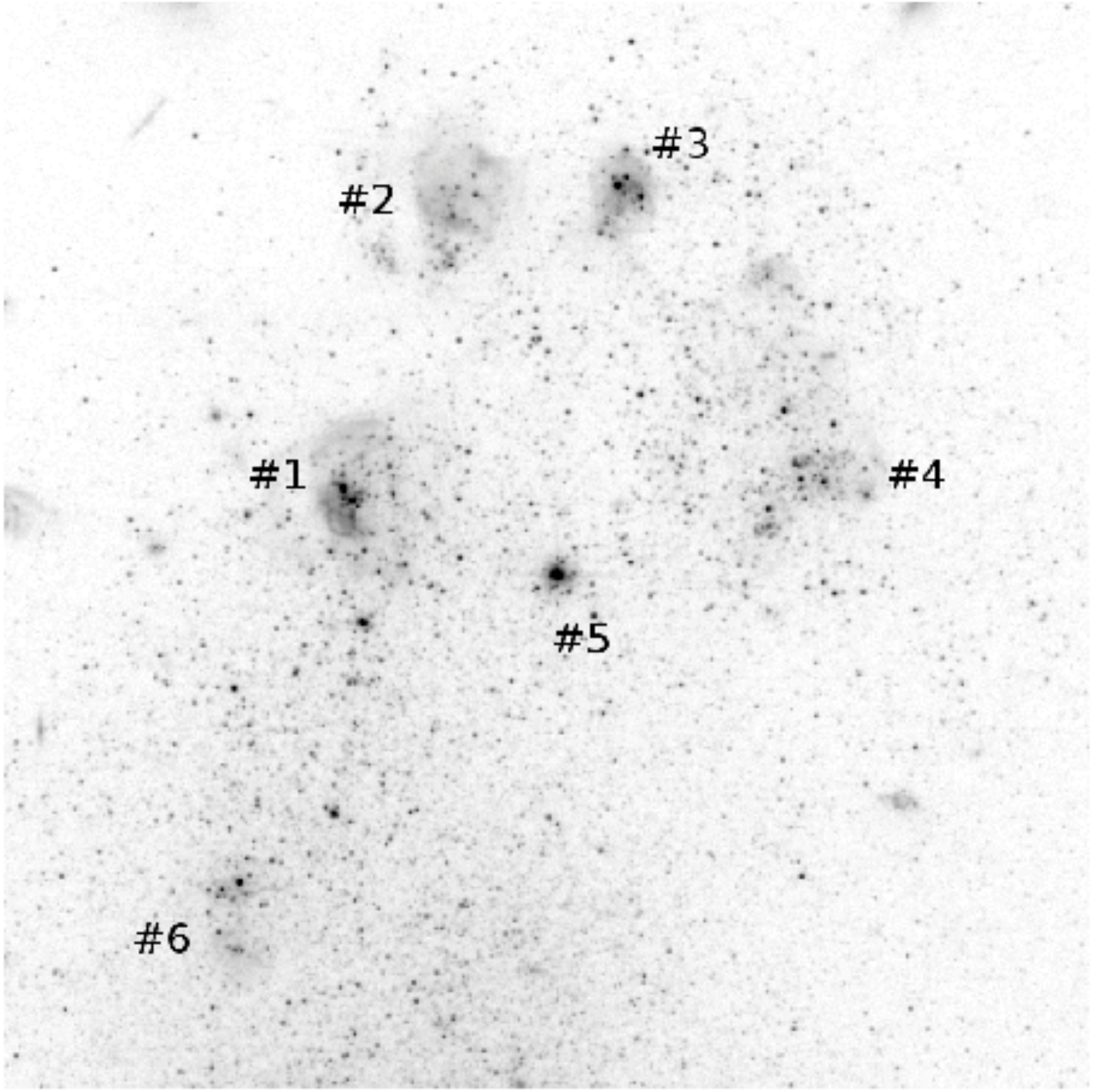}
\includegraphics[angle=0,width=7.5cm, clip=]{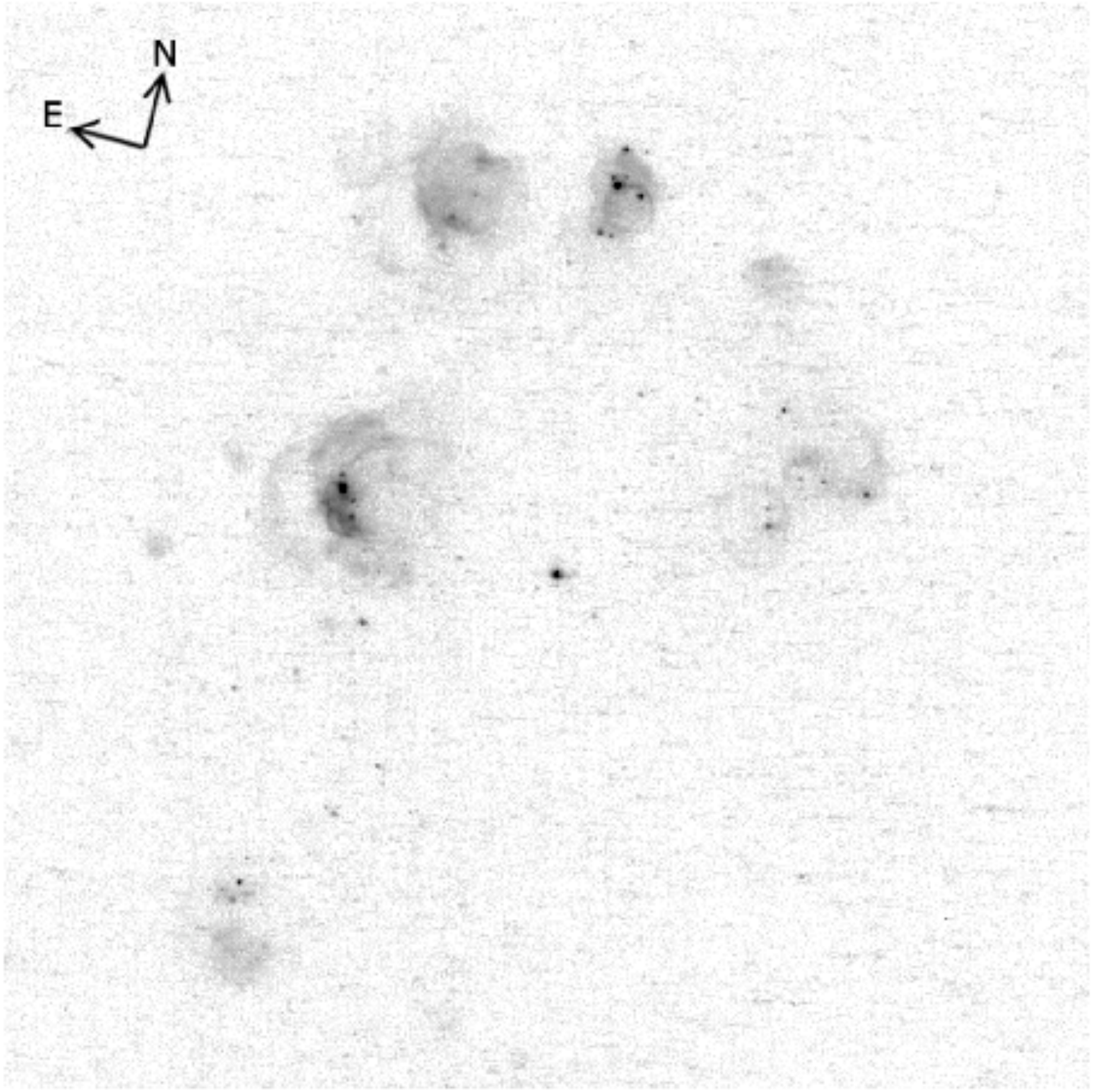}
  \caption{
{\bf Left panel}:
The part of the {\it HST} image of DDO~68 in $V$-band (F606W) centred on the
'Northern Ring' - a collection of regions of active SF, also well seen in
the accompanying image in H$\alpha$. The regions
are situated along the perimeter of a super giant shell (SGS, see text) in
the form of 'ellipsoid'. The minimal ages of the stellar population in
the central part of the SGS as well as the range of \HI\ velocities within
the SGS are used to estimate the time elapsed since the SGS expansion
(see text).
{\bf Right panel}:
The same part of the {\it HST} image of DDO~68 obtained in the narrow band
filter F658N including the line H$\alpha$.
}
	\label{fig:SGS}
 \end{figure*}

The sample appears to be rather diverse with $V-I$ colours in the range
from --0.5 to +1.4 (see Fig.~\ref{fig:CMD50lum}).
Massive O-type stars from the main sequence to supergiants, with
$V-I \le -0.30$ and B-type giants and supergiants (with
$-0.30 < (V-I) < -0.11$) comprise about a half of the most luminous stars.
A-supergiants (with $-0.06 < (V-I) < +0.24$) and F-supergiants
(with $+0.24 < (V-I) < +0.60$)
each comprise $\sim$20~\%.  Finally, K and early M supergiants with
with $+1.3 < (V-I) < +1.6$ comprise $\sim$12~\%.

Since there are many types of variable stars among supergiants, one can hope
to discover such objects with the {\it HST}, similar to the detection of Cepheids
in IZw18 \citep{Aloisi07}.
In fact, many of them can already be monitored to search for optical variability with
the medium-size telescopes installed at the sites with superb visibility.
Therefore, to avoid the future confusions in case of discovery of new
variables in DDO68,
we suggest the alternative name for the DDO68 LBV as DDO68-V1 according to the
common nomenclature for extragalactic resolved variables.

The catalogs of the most luminous stars in the nearby galaxies based on the
HST images are very important and popular among the researchers studying
properties of massive stars. They include both galaxies in the Local
Group (LMC and SMC, M~31, M~33) and more distant representatives of the Local
Volume (e.g., M~101 \citet{Grammer13}, IC~2574). In this paper we are
interested in the most luminous and young stars related to the lowest
metallicity regions in the nearby Universe. They deserve a more detailed
study in order to get insights into
the evolution and feedback effects of massive stars in the early Universe.

\begin{figure*}
  \centering
 \includegraphics[angle=0,width=7.5cm, clip=]{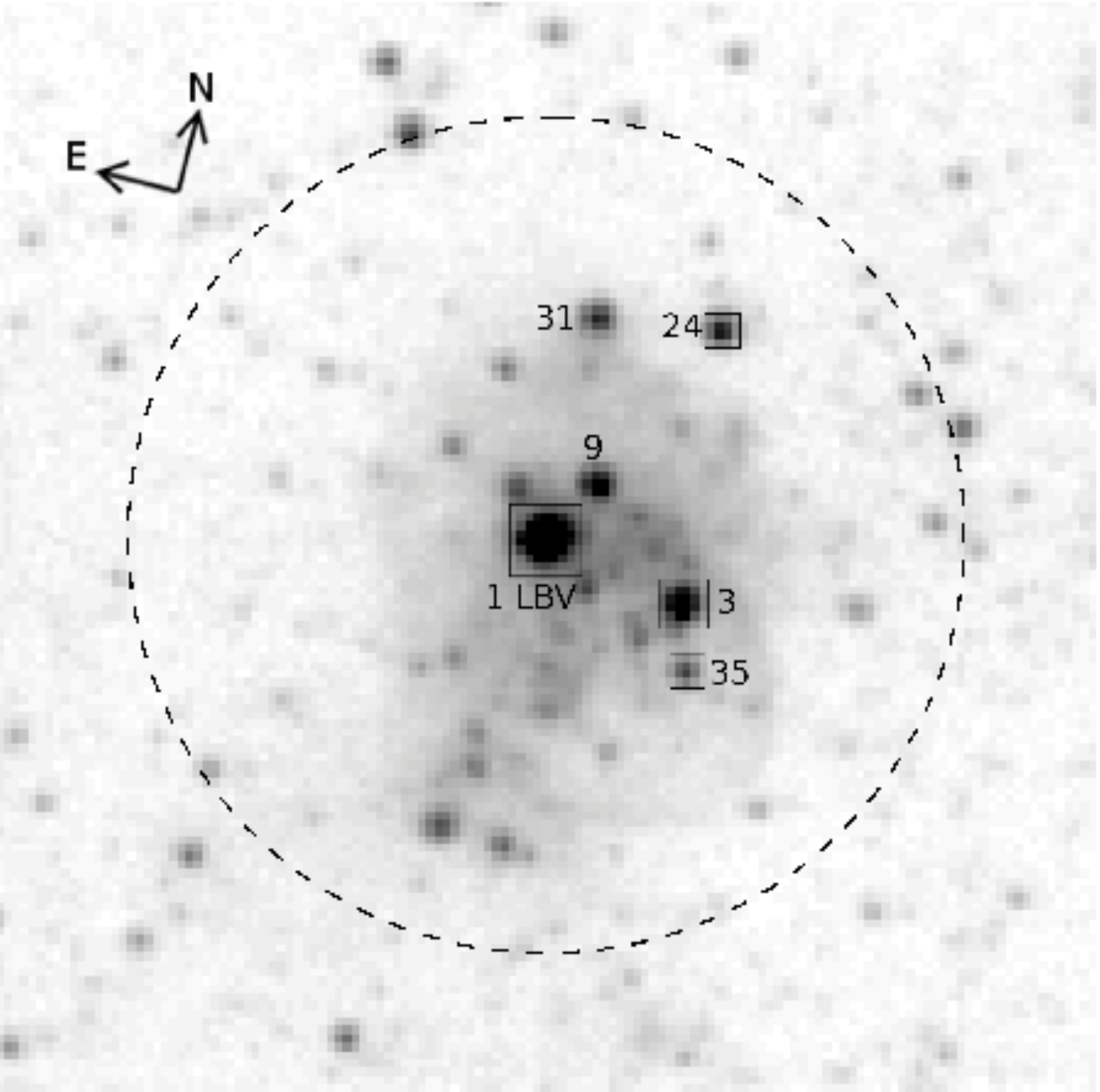}
 \includegraphics[angle=0,width=7.5cm, clip=]{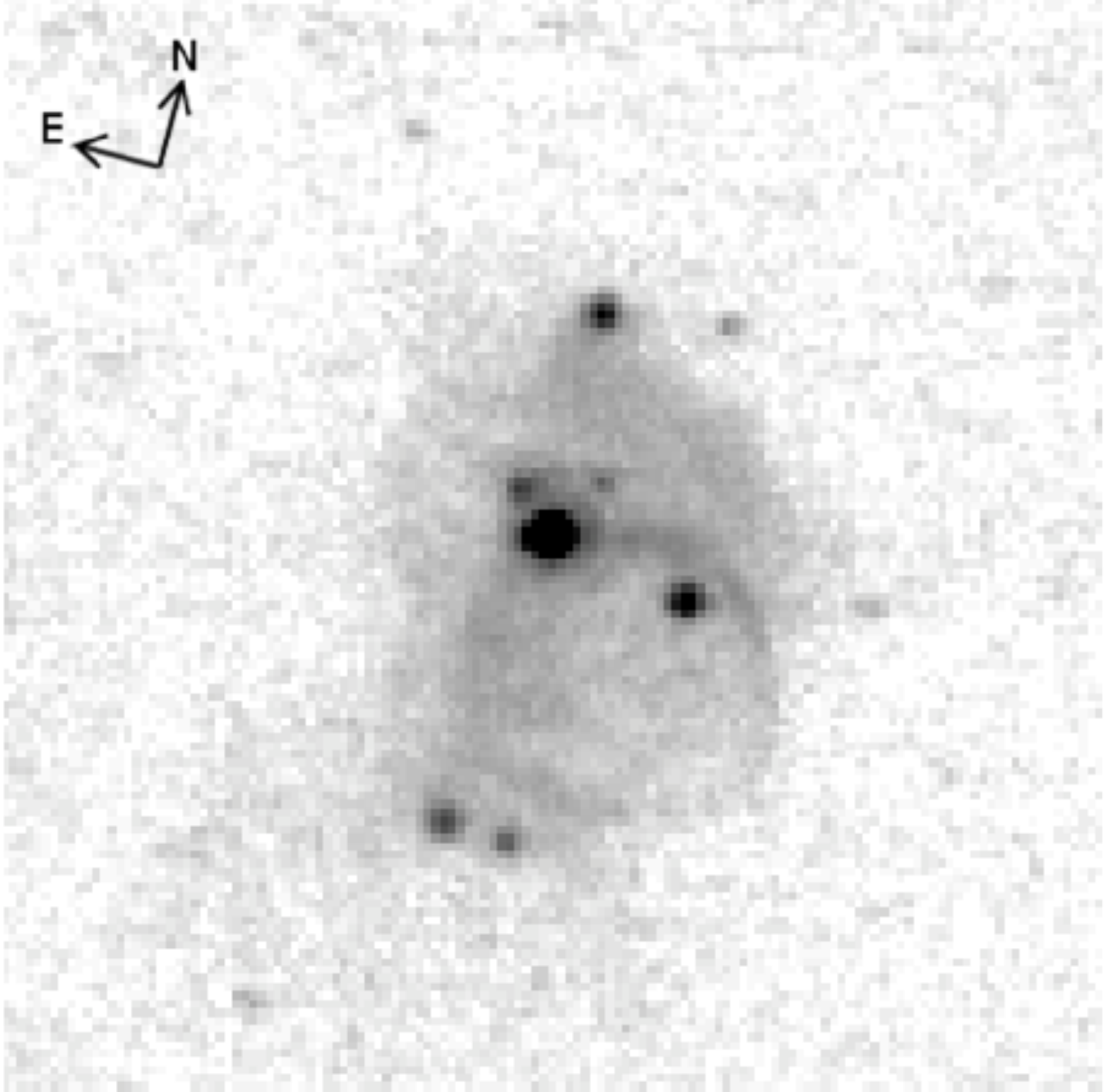}
  \caption{
{\bf Left panel}:
The part of the {\it HST} image of DDO~68 in  $V$ band, centered
  on H{\sc ii} region Knot~3 in the 'Northern Ring' containing the LBV star.
  The LBV, the brightest star-like object on the
image is situated on the rim of the shell-like structure with diameter of
about  $1\farcs 9$, corresponding to $\sim$115~pc. The total extent of
the underlying nebulosity is $\sim$3\arcsec. The dashed-line circle shows
the size of the aperture with $R = 2\farcs 5$. This aperture was used to
measure the full light of Knot~3 in this image and in all other images (SDSS
and BTA) incorporated for constructing lightcurves of Knot~3 and the LBV.
The most luminous stars in
this region are marked by the numbers from Table~\ref{t:lum.list}.
{\bf Right panel}:
 The same part of the {\it HST} image of DDO~68 in the narrow filter F658N
centered at H$\alpha$.  
}
	\label{fig:knot3}
 \end{figure*}

\begin{figure*}
  \centering
 \includegraphics[angle=0,width=5cm, clip=]{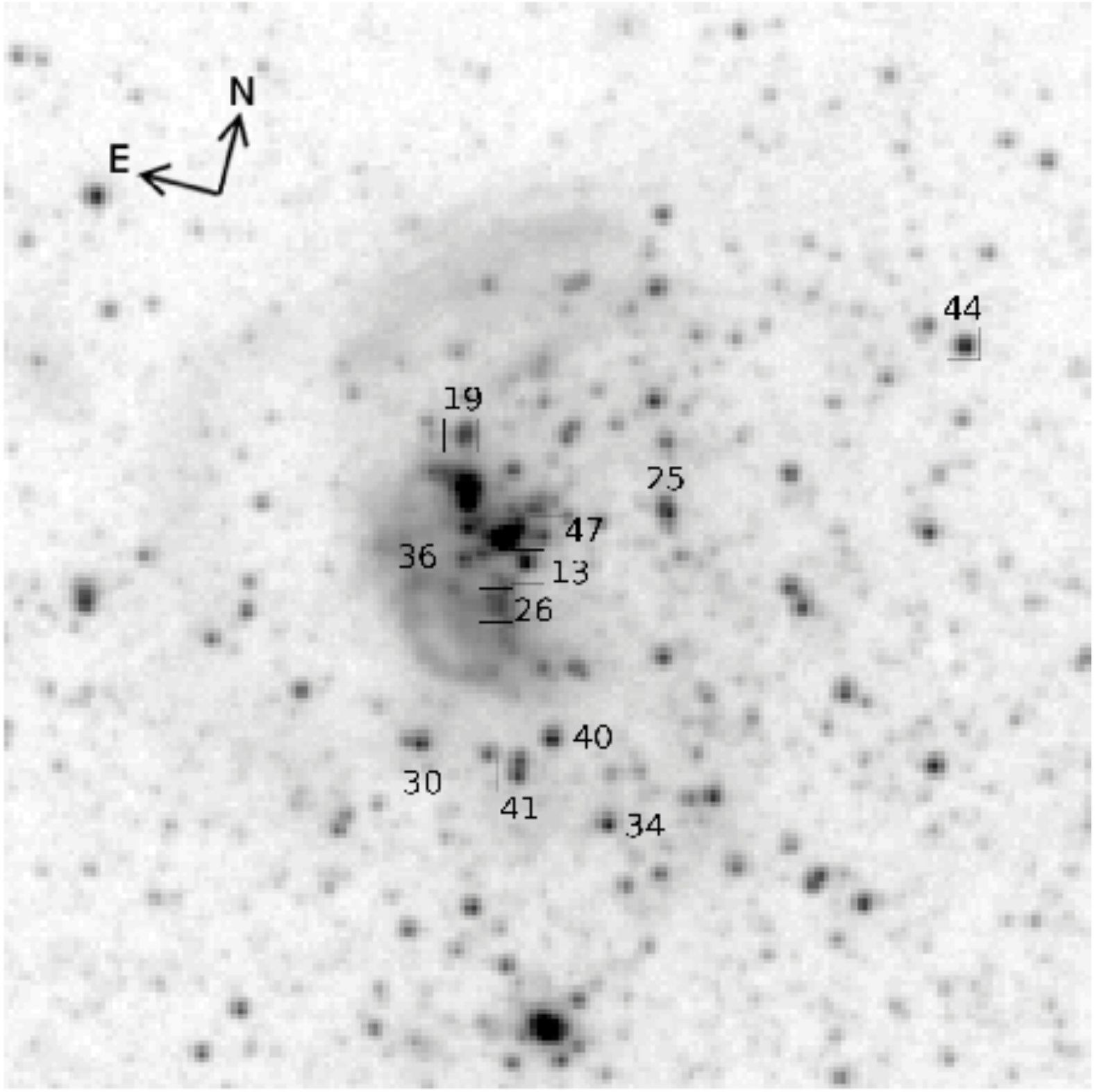}
 \includegraphics[angle=0,width=5cm, clip=]{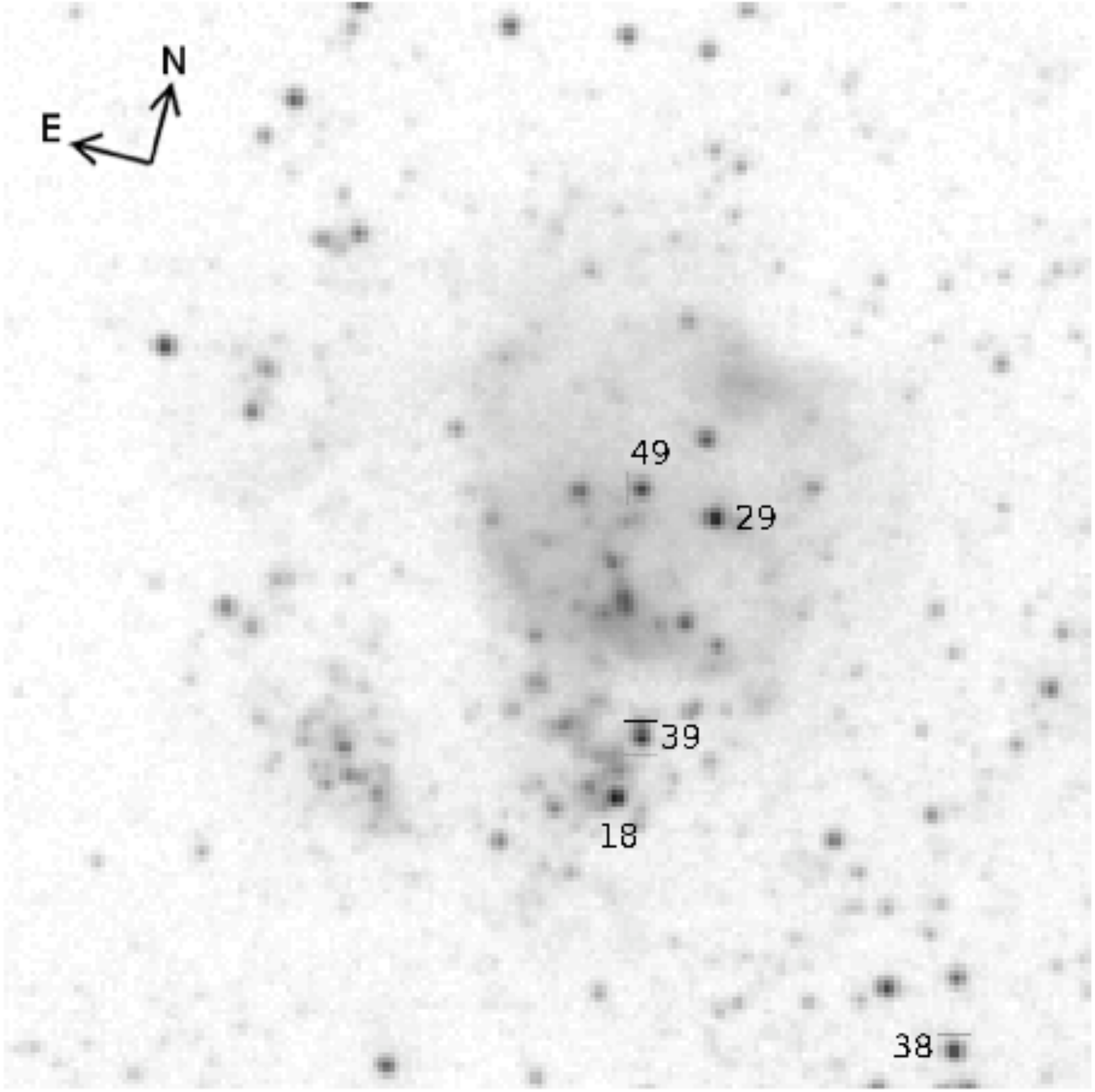}
 \includegraphics[angle=0,width=5cm, clip=]{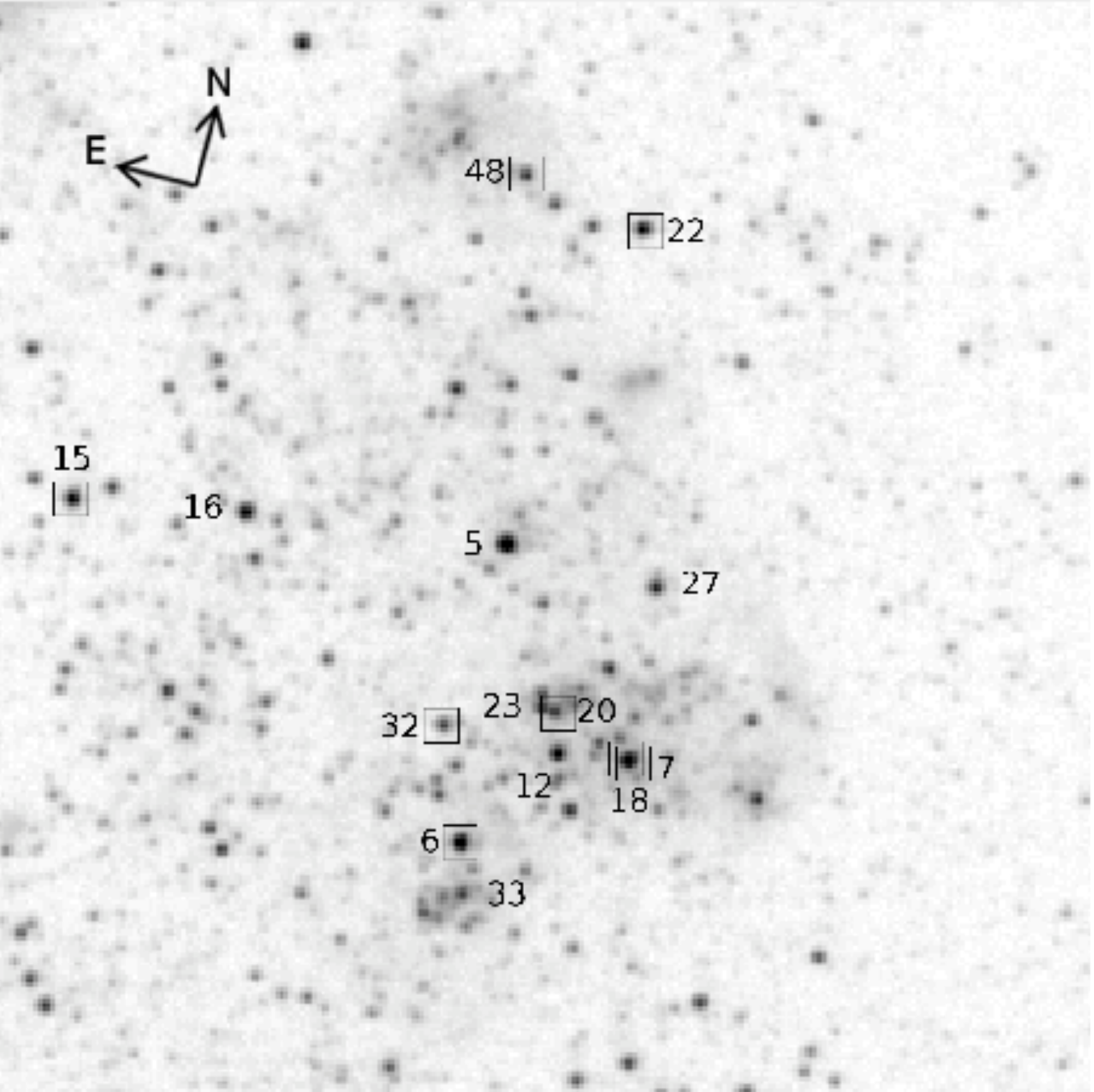}
  \caption{
{\bf Left panel}:
The part of the HST image of DDO~68 in $V$ band centered on SF region
  Knot~1 with marked the most luminous stars.
{\bf Middle panel}: Same for SF region Knot~2.
{\bf Right panel}:  Same for SF region Knot~4.
These and other zoomed in images are intended to better identify the
most luminous young stars in the regions with extremely low gas metallicities.
}
	\label{fig:knot1_2_4_V}
 \end{figure*}

\begin{figure*}
  \centering
 \includegraphics[angle=0,width=5cm, clip=]{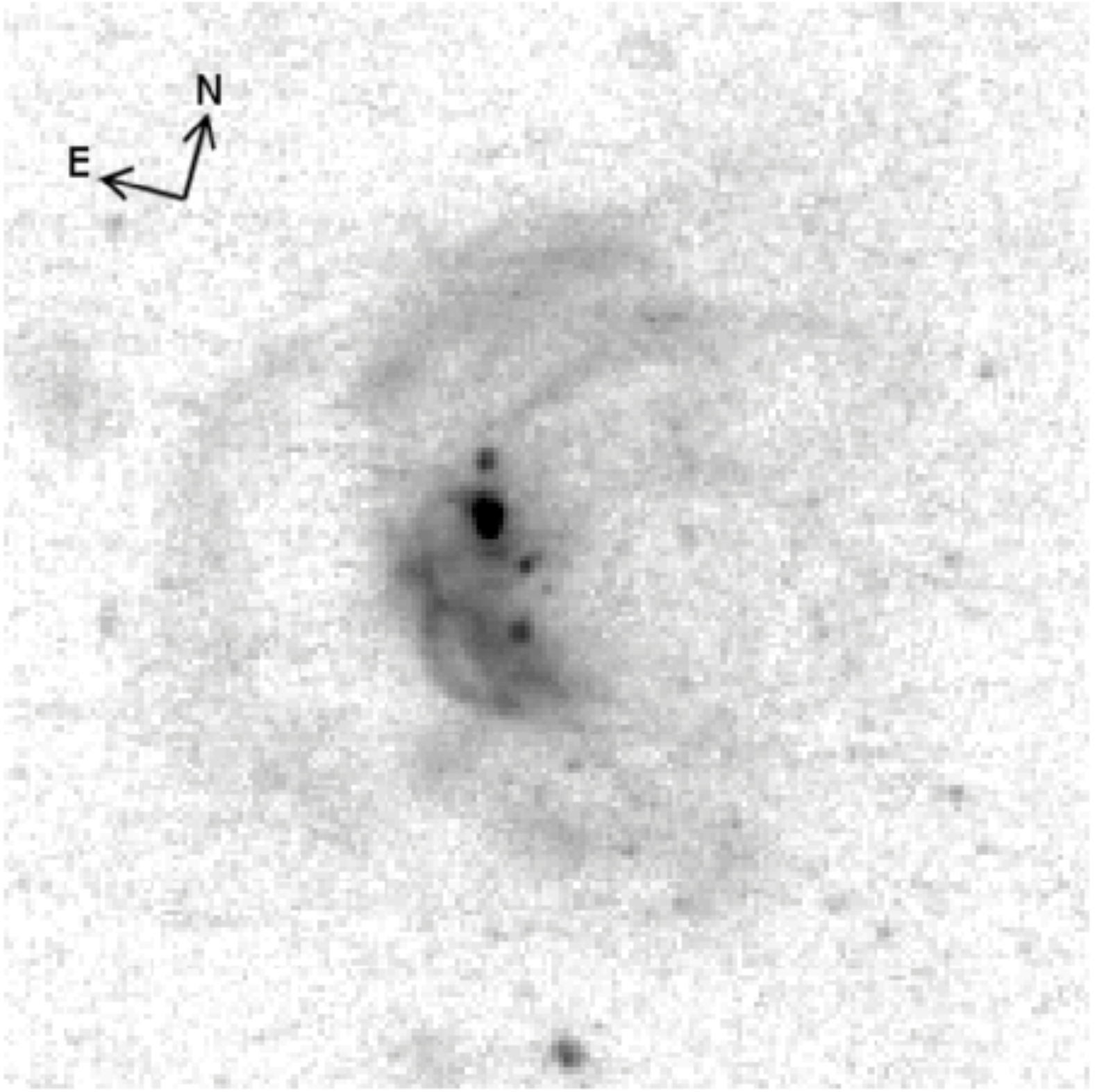}
 \includegraphics[angle=0,width=5cm, clip=]{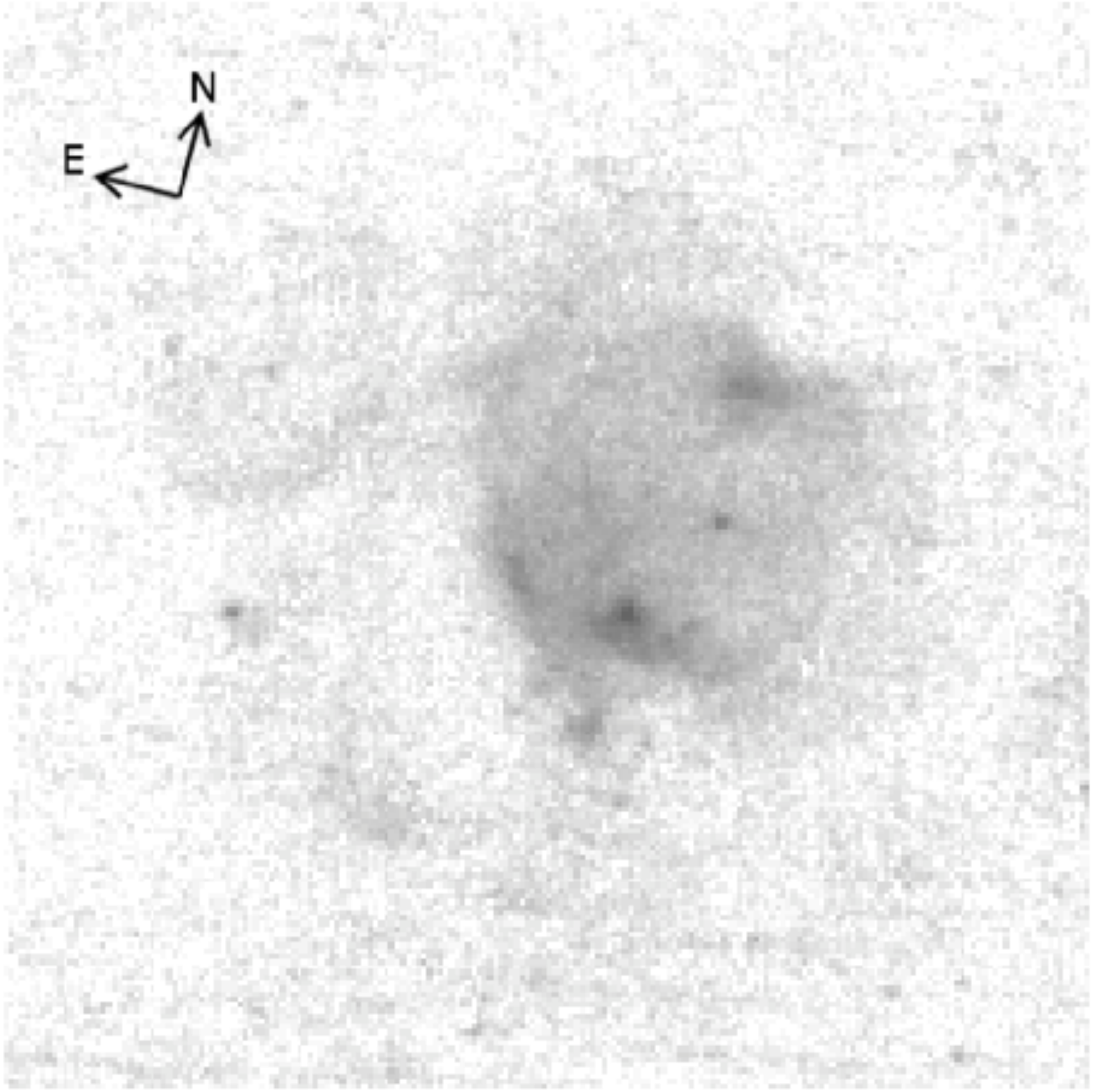}
 \includegraphics[angle=0,width=5cm, clip=]{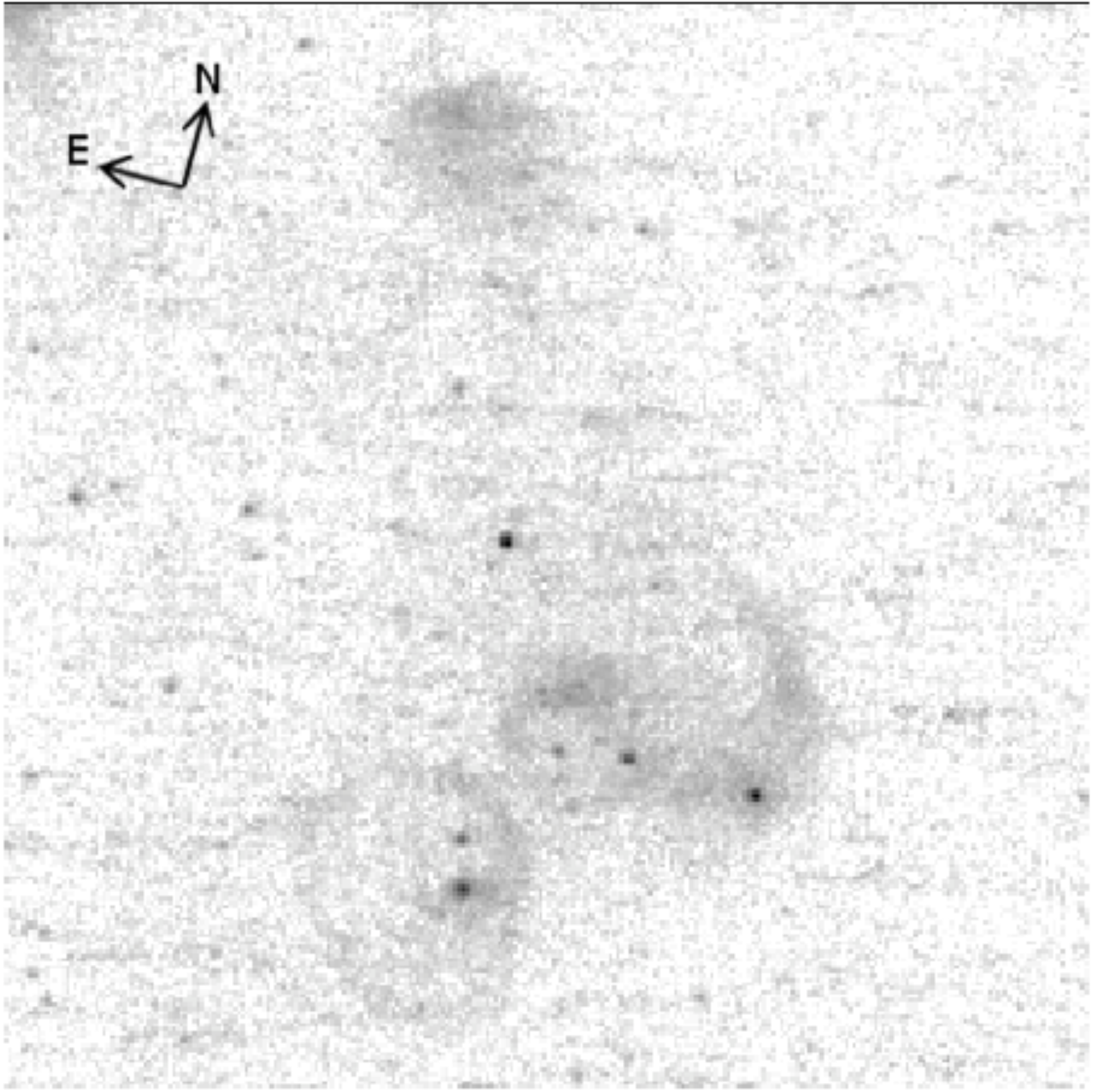}
  \caption{
{\bf Left panel}:
The part of the HST image of DDO~68 in the narrow H$\alpha$ filter centered
 on SF region Knot~1.
{\bf Middle panel}: Same for SF region Knot~2.
{\bf Right panel}:  Same for SF region Knot~4.
}
	\label{fig:knot1_2_4_Ha}
 \end{figure*}

Images of the individual SF regions (Knots~1--7) in both $V$ and
H$\alpha$, shown in Fig.~\ref{fig:knot1_2_4_V} and subsequent ones,
allow one to identify the most luminous
stars in these most metal-poor environments as the attractive targets for
stellar studies with the next generation ELTs.
Besides, careful examination of these images allows one to separate a
subsample of luminous stars, which are substantially isolated, so that the
images with large- and  medium-class  ground-based telescopes
under the seeing of $\beta \sim$0\farcs3 (such as CFHT, Keck, Gemini North)
can be used for regular monitoring and discovery of massive variables of WR
or LBV types with the record low metallicity.

It is worth mentioning that there exist two closer SF dwarf galaxies with very
low metallicities. One is the dIrr galaxy with the gas metallicity
of $\sim Z$\sunn/24  (12+$\log$(O/H)=7.30), residing much closer
than DDO~68. This is UGCA~292 at the distance of $\sim$3.6~Mpc,
with $M_{\rm B}$ = --11.8.
Its SF activity occurs in two main \HII\ regions and the available {\it HST}
data allow one to identify about ten the most luminous stars with $M_{\rm V}$ of
--6.3 to --8.3 ($V$= 21.5 to 19.5) (e.g. \citet{BW11}). These luminous
very low-metallicity stars also deserve further detailed studies, but in
the context of better understanding of the most metal-poor star evolution,
the DDO~68 representatives look at least equally attractive.

Another interesting very small galaxy Leo~P with $Z \sim Z$\sunn/32
(12+$\log$(O/H)=7.17) was recently discovered in the close surrounding of the
Local Group, at D=1.62 Mpc (\citet{LeoP} and references therein).
Its integrated absolute $V$ magnitude $M_{\rm V}$ = --9.3.
Its most luminous star, which ionizes
the only \HII\ region in this galaxy, has $M_{\rm V} \sim$ --4.4 and is probably
a binary of O7 stars. The fainter ones of its most luminous stars are B2--B3 class.

Of somewhat more distant dwarfs in the nearby Universe with the record low
gas metallicity and numerous young massive stellar population, the most
prominent is, of course, the blue compact dwarf IZw18 (12+$\log$(O/H)=7.17)
with a range of distance estimates of 14--15 \citep{IT04} to $\sim$19~Mpc
\citep{Aloisi07,Contreras11}.

We also mention two new Lynx-Cancer void extremely gas-rich LSB dwarfs.
One is SDSS J070623.43+302051.3 (a companion of UGC~3672) at the comparable distance
of $\sim$17~Mpc, with the tentative record-low oxygen abundance of
12+$\log$(O/H)=7.03 \citep{PaperVII,U3672}.
Its star-forming region is much less prominent than those in IZw18. However,
it is comparable or brighter than the SF regions of DDO~68
and with the estimated age of a starburst of $\sim$8~Myr should host late O
and early B stars.

 The second galaxy is also a very gas-rich and record-low metallicity
(12+$\log$(O/H)=7.02) dwarf known as SDSS J094332.35+332657.6
\citep{Hirschauer16}. It was found as an optical counterpart of a new \HI\
source AGC198691
from the blind \HI\ ALFALFA survey \citep{ALFALFA40}. For its kinematic
distance of 10.8~Mpc (assumed to be the same as for its probable companion
UGC~5186), its absolute magnitude  is
$M_{\rm V} \sim -10.68 $. While its HST CMD
is not yet published, the derived H$\beta$ luminosity
$L$(H$\beta$) $\sim$(1--2)$\times 10^{37}$~erg~s$^{-1}$ (based
on the data from \citet{Hirschauer16}) implies that the galaxy can have
several stars of O7V class.

It is also worth mentioning two other extremely metal-poor dwarfs in the
Lynx-Cancer void: SDSS J0926+3343 with 12+$\log$(O/H)=7.12 \citep{J0926} and
SDSS J0812+4836 with 12+$\log$(O/H)=7.26 \citep{IT07} at the estimated
distances of 10.6 and 11.1 Mpc and with absolute magnitudes of
$M_{\rm B} =$ --12.91 and --13.08  \citep{PaperI,PaperIV}, respectively.
They are currently somewhat less active, with the ages of their
starbursts of $\lesssim$ 7~Myr and $\lesssim$ 11~Myr, respectively (as derived
on their EW(H$\beta$) via comparison with the Starburst99 package models
\citep{Starburst99}).
However, the search for their luminous young stars can also reveal
interesting targets for the next generation of giant telescopes.

\newpage

\begin{figure*}
  \centering
 \includegraphics[angle=0,width=5cm, clip=]{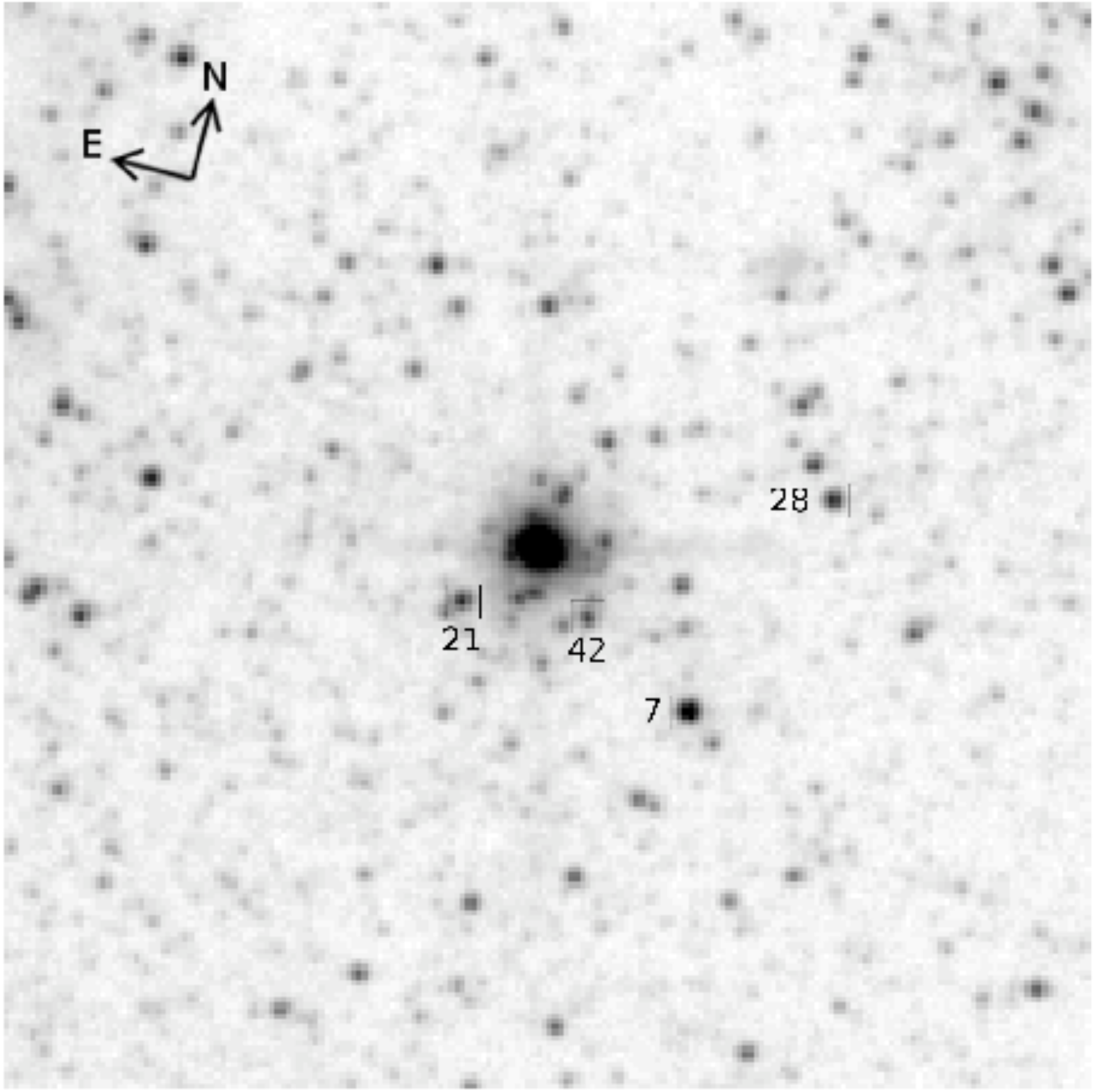}
 \includegraphics[angle=0,width=5cm, clip=]{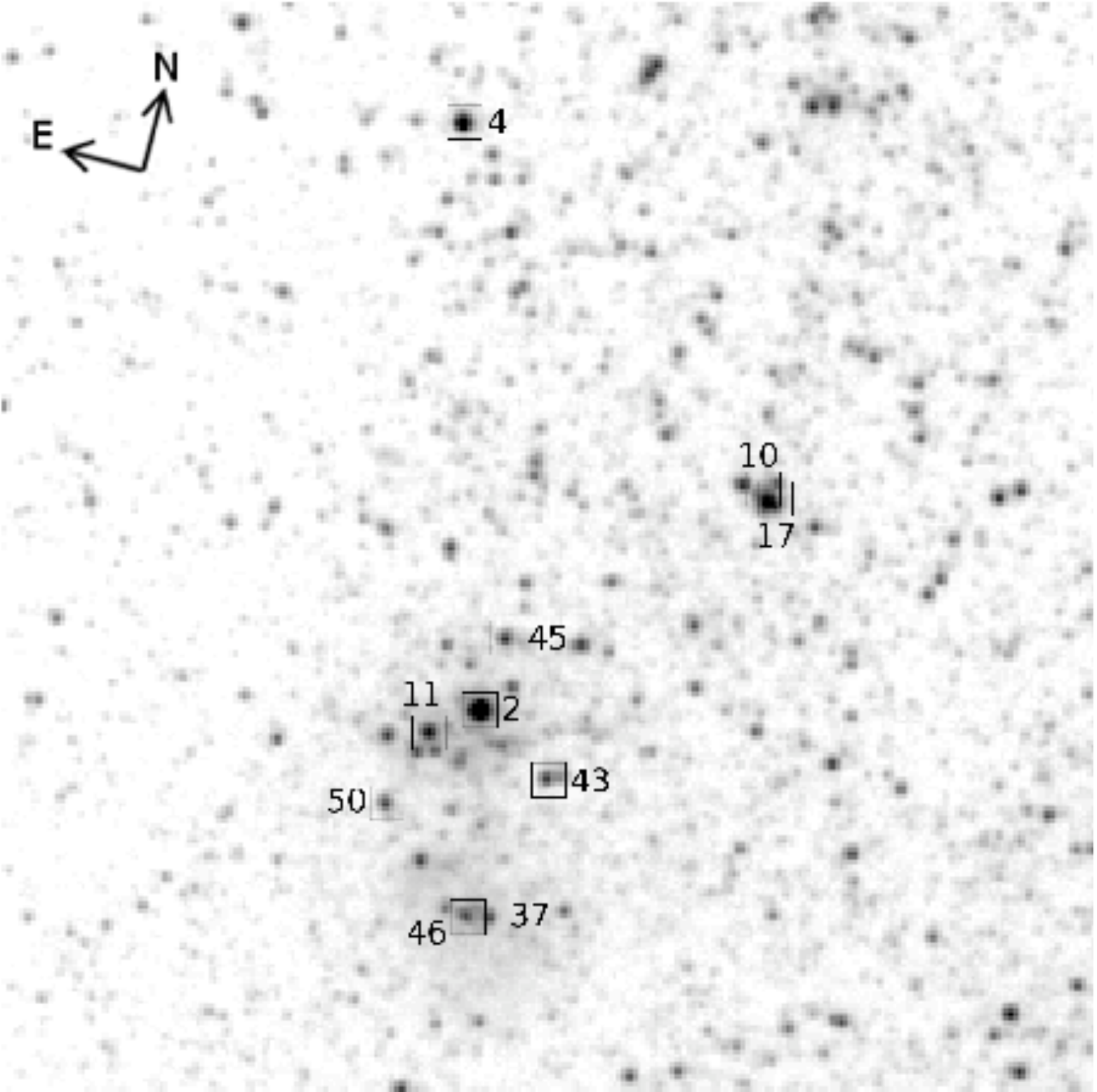}
 \includegraphics[angle=0,width=5cm, clip=]{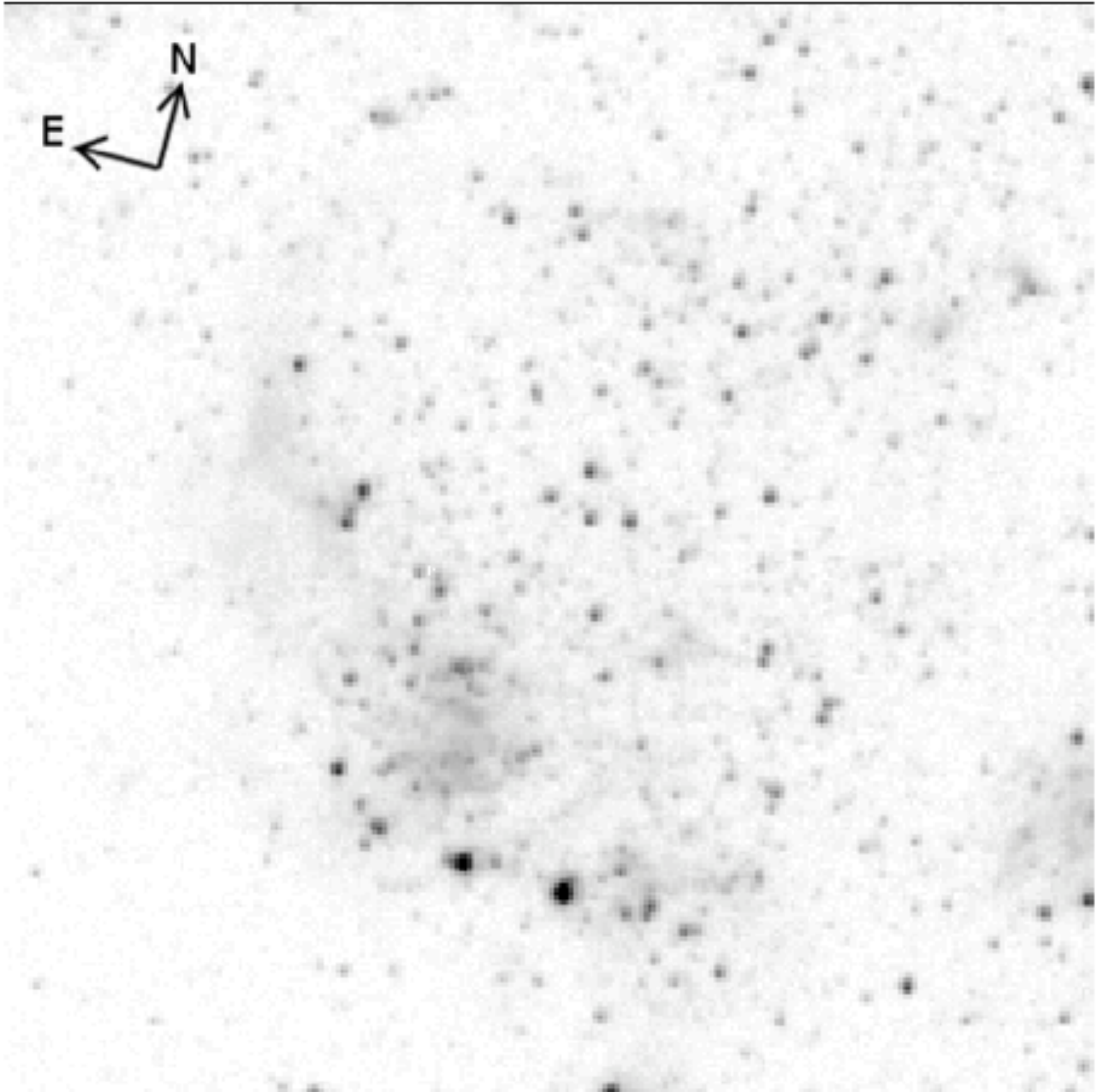}
  \caption{
{\bf Left panel}:
The part of the HST image of DDO~68 in $V$ band, centered
 on Knot~5, a Young Stellar Cluster with marked
  luminous stars in its environments.
{\bf Middle panel}: Same for region of \HII\ region Knot~6.
{\bf Right panel}:  Same for region of Knot~7.
These and other zoomed-in images are intended to better identify the
most luminous young stars in the regions with extremely low gas metallicities.
}
	\label{fig:knot5_6_7_V}
 \end{figure*}

\begin{figure*}
  \centering
 \includegraphics[angle=0,width=5cm, clip=]{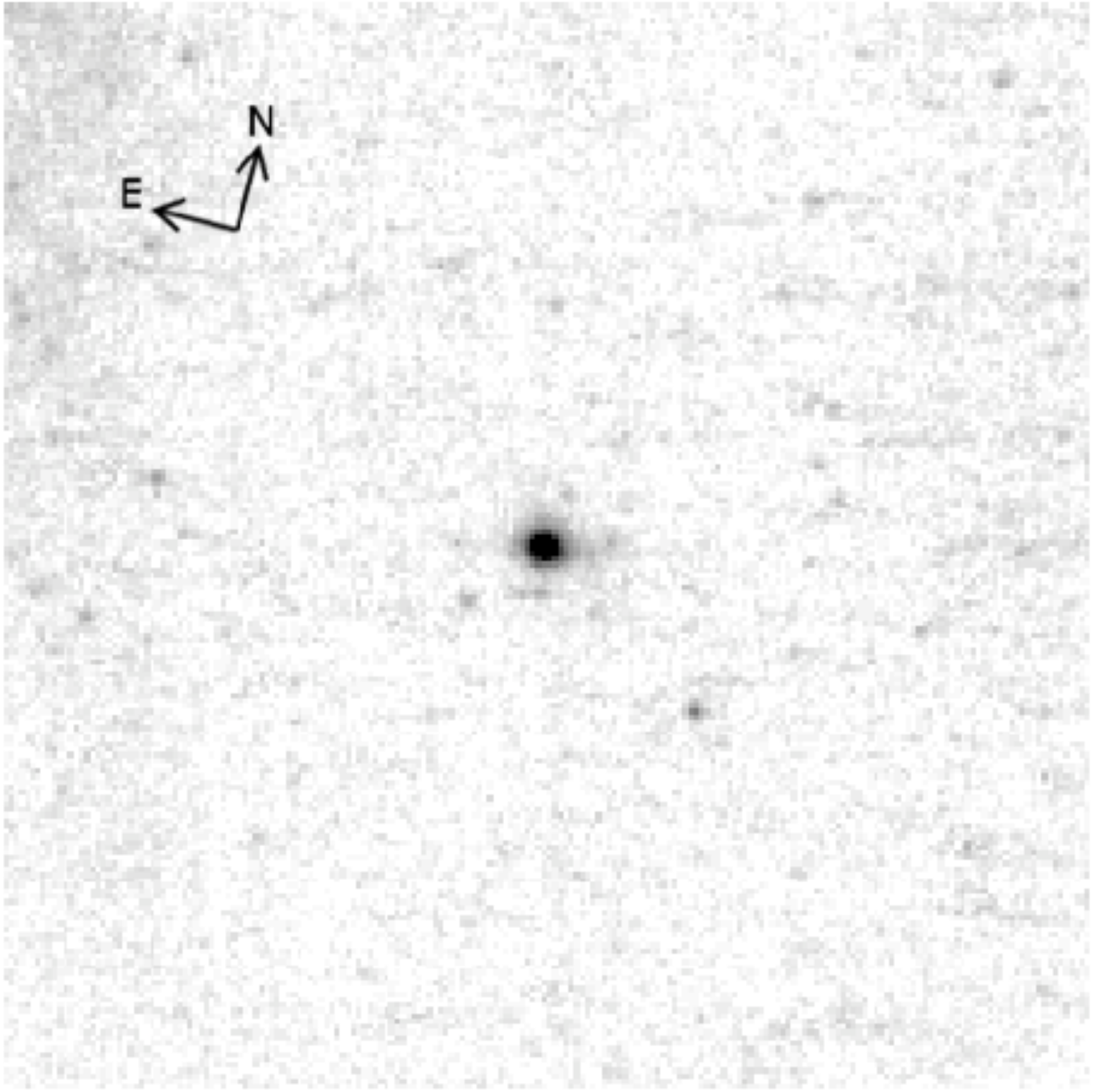}
 \includegraphics[angle=0,width=5cm, clip=]{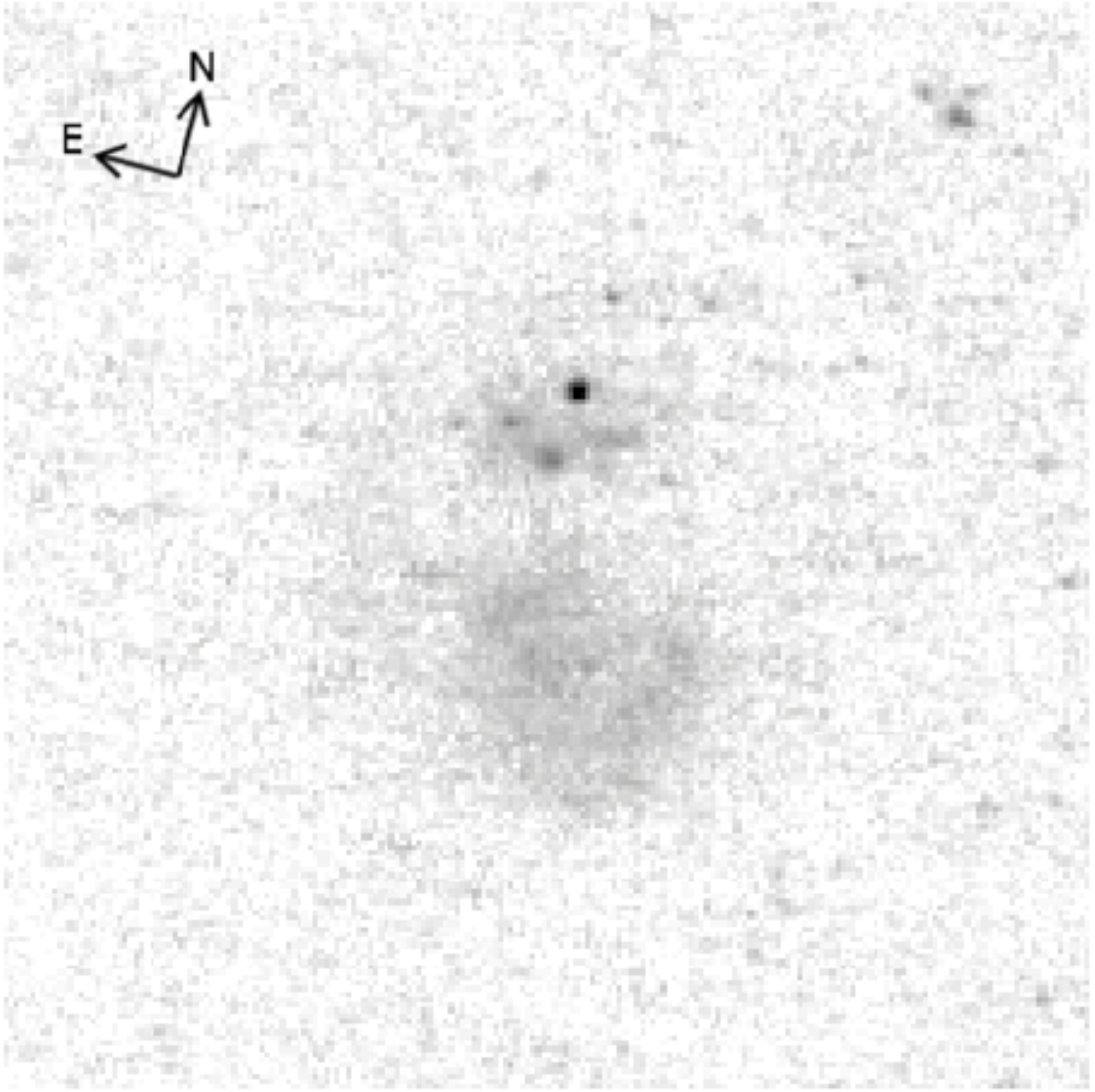}
 \includegraphics[angle=0,width=5cm, clip=]{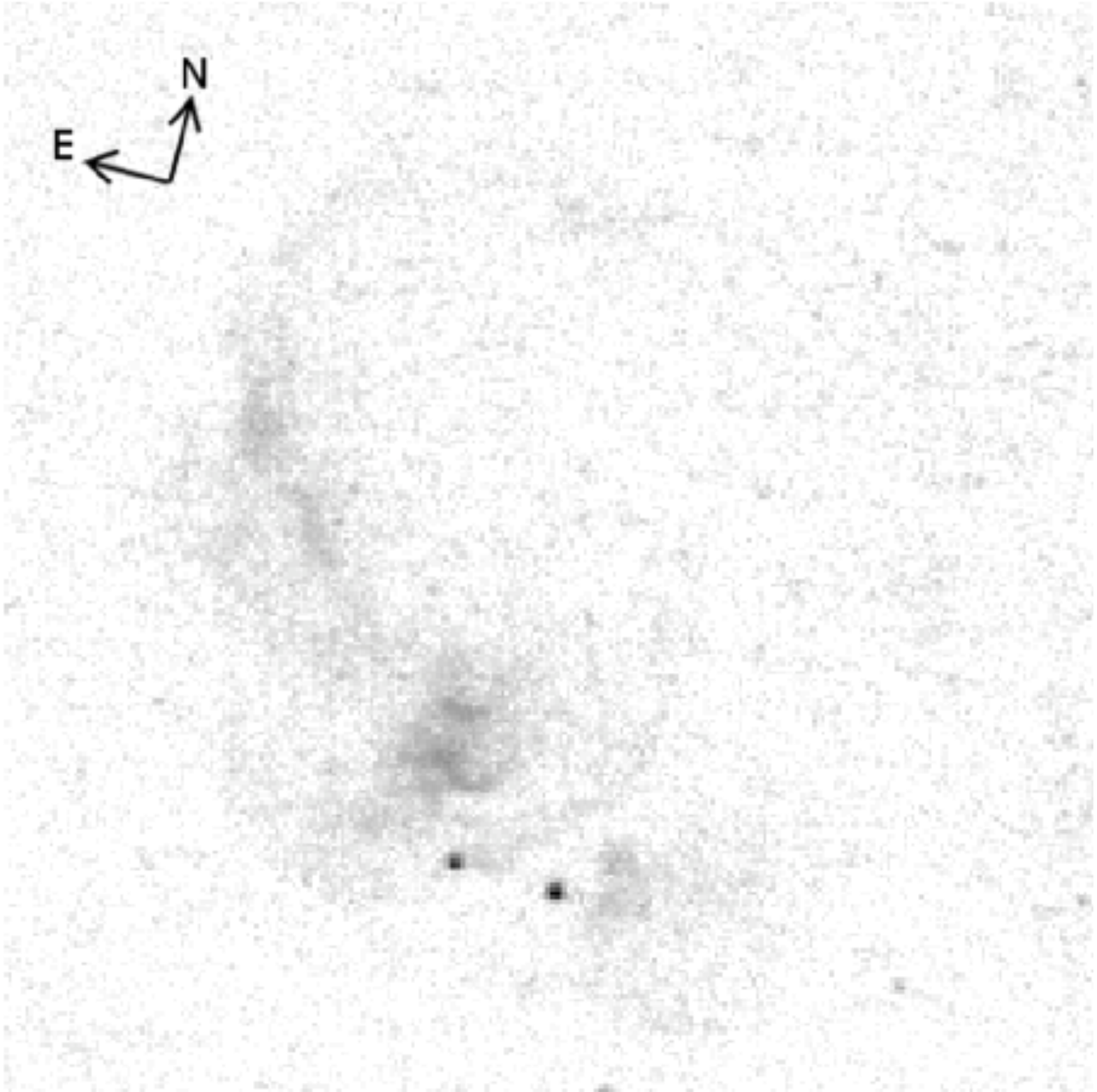}
  \caption{
{\bf Left panel}:
The part of the HST image of DDO~68 in narrow H$\alpha$ filter centered
 on Knot~5, a Young Stellar Cluster.
{\bf Middle panel}: Same for region of Knot~6.
{\bf Right panel}:  Same for region of Knot~7.
}
	\label{fig:knot5_6_7_Ha}
 \end{figure*}

\section[]{DISCUSSION}
\label{sec:dis}

\subsection{The LBV variability in the context}

The typical LBV variability includes variations of 0.1$^m$--0.2$^m$ on short
($\sim$1~day -- weeks) timescale and the irregular variations of up to
$\sim2^m$ amplitude on a timescale of years (called also 'normal eruptions').
During these variations, occurring at 'constant' bolometric luminosity, their
spectral type usually varies from early B supergiant at minimums to late B
or early A near maxima (e.g., \citet{HD94}; \citet{Drissen97,Drissen01};
\citet{Crowther04}; \citet{Massey07}; \citet{Walborn08}).
The largest light variations, with brightening by more than 3$^m$, are
related to 'giant eruptions', in which the mass-loss rate increases by orders
of magnitude and bolometric luminosity increases as well. Such events are
extremely rare among known LBVs. In our Galaxy, they are registered only
for two classic LBVs: $\eta$~Car and P~Cyg. Among LBVs in nearby galaxies
only a handful of giant eruptions were discovered.

Despite the fact that the amount of photometric and spectroscopic data for DDO68-V1 is
very modest, it is useful to make the preliminary comparison with other
well studied LBVs with giant eruptions. One of them is NGC2363-V1 at
$D \sim$3.4~Mpc, for which the giant eruption had taken place since the end of 1993.
Starting from $M_{\rm V} \sim$ --6.5~mag, it reached the maximum of lightcurve
at $M_{\rm V} \sim$ --10.5~mag at the end of 1997 and then gradually faded by
$\Delta V \sim$ 0.3~mag during the next
7 years \citep{Drissen97,Drissen01,Petit06}. In course of this fading in $V$,
its UV flux was growing by about 0.6$^m$, so its bolometric luminosity
had increased during this period.

The behaviour of DDO68-V1 (Fig.~\ref{fig:lightcurve}, bottom) shows
substantial differences with respect of the lightcurve of NGC2363-V1.
The whole range of light variations $\Delta V \sim$ 3.6$^m$ ($V \gtrsim$--6.95
--10.58) is quite similar to that of NGC2363-V1. However, while for
NGC2363-V1 we see more or less monotonic luminosity increase during $\sim$5
years, for DDO68-V1  the pre-eruption activity was rather strong. In
particular, before the 'deep' minimum at the end of 2004 and the beginning
of 2005, in April 2004 (SDSS image), the LBV was in an intermediate phase,
$\sim$1.5$^m$ fainter than for the eruption peak $V$-magnitude. It also
was in an intermediate phase and near the maximum, according to Knot~3 light
photometry in $B$ by  \citet{BW11}, in the period between 1993 and 2001. It is
probable that DDO68-V1 was close to the maximum optical light in 1955 (also
as measured by \citet{BW11}, based on the POSS photoplates).

Due to the poorly documented, but seemingly rather erratic (unpredictable)
and strong variability of DDO68-V1, it can be a fainter analog of an unusual
('persistent') extragalactic LBV known as UGC~2773-OT \citep{Smith2016}.
While this is a very tempting option, it would be crucial to conduct its regular multiband monitoring
to have a more certain classification of
LBV DDO68-V1.

As \citet{Smith2006} discuss, the series of such giant eruptions in LBVs,
which form several expanding shells, can precede their SN explosions on a
rather short time scale. The most recent example of such
an event is presented in \citet{Vinko15,SNHunt275} and references therein,
where the LBV/SN impostor in galaxy NGC~2770 exploded as SNII within a period of just three
months after its giant eruption. Thus, if the erratic variability of this
LBV will be confirmed,
DDO68-V1 can be a good candidate to
study progenitors of SN with the lowest metallicity in the nearby Universe.

The only issue that remains open in qualifying the recent DDO68-V1
large amplitude variations as giant eruptions is its unique low metallicity.
The constant bolometric luminosity, characteristic of variations with
$\Delta V \lesssim$ 2$^{m}$ ('normal eruptions') is the empiric fact derived
from the studies of much more metal-rich LBVs. Since this is not yet well
understood and based on models, in principle one can suggest that for the
most metal-poor LBV representatives, like DDO68-V1, the range of large
$V$-band variations with the constant bolometric luminosity can be wider.
Therefore, to firmly confirm DDO68-V1 giant eruption(s), it is crucial
to extend the monitoring of this LBV to the wider wavelength range to see
how its spectral energy distribution (SED) and bolometric luminosity
vary during the large visual brightening.

\subsection{DDO~68-V1 spectral variations in comparison to the known LBVs}
\label{ssec:spectra_compar}

The described relation of \HeI(4471) flux with the LBV $V$-band luminosity
differs from that found in the systematic study of LBV V532 in the Local
Group galaxy Messier 33 \citep{Sholukhova11,Fabrika05}. They found that
during eleven years variations, $F$(\HeI~4471) and $F$(\HeI~5876) correlated
with V532 optical luminosity
variations for the total amplitude of $\Delta B =$1.6$^m$. The total range
of light variations in this LBV is known to be $\sim$2.5$^m$,
with the minimum level of $M_{\rm V} \sim$ --6.1.

On the other hand, the fluxes of \HeI\ lines in DDO68-V1 show the behaviour
somewhat similar to that of the well-studied LBV NGC~2363-V1 \citep{Petit06}.
Due to the complicated environment, the spectra of this LBV was possible to
observe only with the {\it HST}. The collection of its UV-optical spectra was
obtained during seven years, starting from the maximum of the 'giant eruption'
in 1997.
$V$ band flux after $\sim$1~year fluctuations dropped gradually during this
period by $\sim$0.3$^m$
while the UV continuum varied in the counterphase and increased by 0.6$^m$.
Fluxes of \HeI\ lines (and in particular, \HeI~4471 and \HeI~5876) varied in the
opposite sense with respect of $V$ band changes. They were faint (\HeI~5876)
or in absorption (\HeI~4471) near the maximum of the $V$ lightcurve. Then they
were strongly rising to the epoch of the conditional (local) 'minimum':
\HeI~5876 flux
increased by factor 5--6, while \HeI~4471 as well as other faint \HeI\ lines
became well detected in emission.
At the same time, the fluxes of Balmer lines gradually decreased during this
LBV evolution, with F(H$\alpha$) drop by a factor of 4. Without more details,
we can summarize that in this LBV the fluxes of \HeI\ lines grow several-fold
when the LBV gets dimmer in $V$-band after the giant eruption,
while the fluxes of Balmer emission lines drop by comparable amount.

The comparison of Balmer and \HeI\ line flux variations in DDO68-V1 with the
well-studied LBVs described above leads one to the following conclusions.
For its current rather small statistics, fluxes of \HeI\ lines in DDO68-V1
changed more or less similar to those of NGC~2363-V1 near its maximum light
phase.
For DDO68-V1  Balmer line flux variations, the situation is somewhat unusual.
While for H$\alpha$ we have only two estimates, we use in the following only
data for the flux of H$\beta$. This appeared about the same (and about the
average over the whole set) for dates separated by 76 days, when the LBV was
in the maximum and in the phase of 2$^m$ fainter. Furthermore, in 10 months
after the maximum when the LBV $V$-band luminosity dropped by $\sim$50~\%,
the flux F(H$\beta$) had increased by the same $\sim$50~\%.

It is difficult to find an evident cause of such Balmer line behaviour. One
of the possible reasons
could be the aforementioned unusual erratic activity of DDO68-V1 during
several last decades, which forms the sequential (interacting?) shells/winds.
In this case the simple correlations visible in NGC~2363-V1 giant eruption
phase may not work.

\section{Summary}
\label{sec:summ}

We summarize the presented results and the discussion above and draw the
following conclusions:

\begin{enumerate}
\item
One of the brightest stars on the HST images of the void galaxy DDO~68
is identified with the DDO~68 Luminous Blue Variable
(LBV) discovered in January 2008 (also DDO68-V1). We present the new BTA
photometry of \HII\ region Knot~3 (containing the LBV) complemented with the
photometry from the resolved HST images and from the SDSS image archive.
The HST images allow one to measure separately the light from the LBV
and the underlying \HII\ region and to use the latter in the analysis of the
ground-based data of this complex.
Along with the analysis of new and old spectra of this region at several
epochs, this data, for the first time, allows one to determine the reliable amplitude
of the LBV lightcurve. All available data suggest that the LBV $V$ magnitude
varied during the last decade in the range of $\sim$20.0 to fainter than
23.6 mag. This corresponds to $M_{\rm V}$ range of fainter than --6.95 to
--10.58. If one combines our lightcurve since 2004 with the earlier estimates
of Knot~3 magnitudes from \citet{BW11} since 1955, the high LBV state
assumed to relate to giant eruptions, was also reached  in 1955 and in 1999.
If photometric behaviour of the most metal-poor LBVs is similar to more
typical LBVs, the DDO68-V1 light variations during the last
20 (60) years suggest that it probably underwent the giant eruptions.
Multiwavelngth monitoring would be useful to prove the substantial increase
of the LBV bolometric luminosity.
\item
To the spectra of the LBV obtained in January, February and March 2008, we
added one more obtained at BTA in January 2009. Despite the very modest
amounts of spectral data, we try to compare them with known LBVs in the giant
eruption phase. All spectra of DDO68-V1 are analyzed together to establish
some common or atypical features, which are important in the context of very
low metallicity of the LBV progenitor. We confirm the P-Cyg profiles in the
Balmer hydrogen lines and the terminal wind velocity of $\sim$800~\kms.
We find a substantial decrease of line fluxes of \HeI$\lambda$4471 and
$\lambda$5876~\AA\
while the LBV $V$-band luminosity increased by $1.5^m-2^m$, in general
consistent with \HeI\ line flux changes in giant eruption phase in other LBVs.
However, fluxes of Balmer emission lines do not show clear correlation with
the LBV luminosity. This hints to an unusual LBV type. However, due to
the small volume of statistical data, the results of additional spectral observations of
DDO68-V1 will be crucial.
\item
The prominent DDO~68 'Northern ring' of SF regions is examined using the
HST data, the BTA FPI H$\alpha$ data and the GMRT \HI-maps. All combined data
are consistent with the scenario of the powerful SF episode near the
'Northern ring' center lasting for tens Myr since $\sim$30~Myr ago.
Cumulated action of multiple SNe resulted in a supergiant shell (SGS) with
the observed diameter of 1.1~kpc.
The instabilities in the SGS gas layer have induced the second
generation Star Forming complexes seen on the ground-based images as
six regions along the SGS perimeter. On the HST images, these SF regions
show complex substructure including multiple luminous stars and multiple
arc and ring structures with radii of $\sim$50 to $\sim$200 pc.
They are related with SF episodes in these complexes. With $V \sim$
20~\kms, their kinematical ages fall in the range of $\sim$1--7~Myr.
Accounting for the delay of SN explosions of 3.5~Myr since the SF onset,
they are consistent with the independent estimates of young SF region
ages (3--7~Myr).
For majority of these secondary SF complexes, in the vicinity of the
brightest (youngest) arcs, one finds compact SF knots with sizes close
to the resolution limit ($\lesssim$12 pc). They may represent the next
generation of the induced SF in the 'Northern ring'.
The LBV DDO68-V1 looks to belong to this generation since it is situated
very close to the related H$\alpha$ arc.
\item
Using the HST archive F606W and F814W images of DDO~68, we compile the list
of the most luminous stars in the regions of the latest active SF, in which
the measured ionized {\it gas} metallicity is at the level of $Z$
$\sim Z$\sunn/35. The total number of such stars brighter than $M_{\rm V}
< -6.0$ ($20.05^{m} < V < 24.56^{m}$) is about half a hundred.  These massive
diverse young stars (from O to M supergiants), by their origin in the
aforementioned 
star-forming regions, are the most metal-poor massive stars in the nearby
Universe. They will be accessible for the detailed studies with the next
generation of giant ground-based (like E-ELT and TMT) and space telescopes
(JWST). Therefore, they are among the best candidates for comparison
with predictions of the modern models of massive star evolution at very
low metallicities. As preparation for detailed studies, a fraction of
these stars can be monitored for optical variability with the existing medium-size
ground-based telescopes located in the sites with the superb visibility.
\end{enumerate}

\section*{Acknowledgements}

The results for the LBV,  H$\alpha$ shells and sequential induced episodes
of star formation are obtained with the support from RSCF grant No.
14-50-0043 to SAP and YAP.
AVM is grateful for the financial support via the grant
MD3623.2015.2 from the President of the Russian Federation. We thank
the anonymous referee for many useful advises, questions and suggestions
which allowed us to improve the paper.
SAP thanks \mbox{Y.I.~Izotov} for stimulating conversations during
the very early stages of this work. We are pleased to thank J.~Chengalur and
E.~Egorova for help with GMRT high resolution \HI-map and S.~Fabrika
and O.~Sholukhova for consultation on spectral variability of LBVs.
Observations at the 6-m telescope BTA are supported by funding from the
Russian Federation Ministry of Education and Science (agreement
No~14.619.21.0004, project identification RFMEFI61914X0004).
Some of the results in this article are based on observations made with the
NASA/ESA Hubble Space Telescope, and
obtained from the Hubble Legacy Archive, which is a collaboration between
the Space Telescope Science Institute (STScI/NASA), the European Space Agency
(ST-ECF/ESAC/ESA) and the Canadian Astronomy Data Centre (CADC/NRC/CSA).
The {\it HST} observations presented in this work are associated with program
11578 (PI A.Aloisi). We acknowledge the use of the SDSS DR7 database.
Funding for the Sloan Digital Sky Survey (SDSS) has been provided by the
Alfred P. Sloan Foundation, the Participating Institutions, the National
Aeronautics and Space Administration, the National Science Foundation,
the U.S. Department of Energy, the Japanese Monbukagakusho, and the Max
Planck Society. The SDSS Web site is http://www.sdss.org/.
The SDSS is managed by the Astrophysical Research Consortium (ARC) for the
Participating Institutions.


\bsp

\label{lastpage}

\end{document}